\PassOptionsToPackage{unicode,pdfa}{hyperref}
\PassOptionsToPackage{hyphens}{url}
\PassOptionsToPackage{dvipsnames,svgnames,x11names}{xcolor}
\documentclass[
  a4paper,
]{article}
\usepackage{amsmath,amssymb}
\usepackage{iftex}

  \usepackage[T1]{fontenc}
  \usepackage[utf8]{inputenc}
  \usepackage{textcomp} % provide euro and other symbols

\usepackage{lmodern}
\ifPDFTeX\else
\fi
\IfFileExists{upquote.sty}{\usepackage{upquote}}{}
\IfFileExists{microtype.sty}{% use microtype if available
  \usepackage[]{microtype}
  \UseMicrotypeSet[protrusion]{basicmath} % disable protrusion for tt fonts
}{}
\makeatletter
\@ifundefined{KOMAClassName}{% if non-KOMA class
  \IfFileExists{parskip.sty}{%
    \usepackage{parskip}
  }{% else
    \setlength{\parindent}{0pt}
    \setlength{\parskip}{6pt plus 2pt minus 1pt}}
}{% if KOMA class
  \KOMAoptions{parskip=half}}
\makeatother
\usepackage{xcolor}
\usepackage[top=0.5cm,right=2.5cm,bottom=2.5cm,left=2.5cm]{geometry}
\usepackage{longtable,booktabs,array}
\usepackage{calc} % for calculating minipage widths
\usepackage{etoolbox}
\makeatletter
\patchcmd\longtable{\par}{\if@noskipsec\mbox{}\fi\par}{}{}
\makeatother
\IfFileExists{footnotehyper.sty}{\usepackage{footnotehyper}}{\usepackage{footnote}}
\makesavenoteenv{longtable}
\providecommand{\tightlist}{%
  \setlength{\itemsep}{0pt}\setlength{\parskip}{0pt}}
\input{preamble.sty}
\usepackage{pdflscape}
\usepackage{multirow}
\usepackage{array}
\usepackage{longtable}
\usepackage[inkscapepath=svg-inkscape]{svg}
\newcommand{\blandscape}{\begin{landscape}}
\newcommand{\elandscape}{\end{landscape}}
\newcommand{\fonticon}[2]{\includesvg[height=1.5ex]{{./fonts/#1}}\nobreakspace{}#2}
\newcommand*{\DOCBB}{\fonticon{sitemap}{DOCBB}}
\newcommand*{\DIST}{\fonticon{boxes-packing}{DIST}}
\newcommand*{\SWLC}{\fonticon{arrows-spin}{SWLC}}
\newcommand*{\SWREPOS}{\fonticon{code-pull-request}{SWREPOS}}
\newcommand*{\MOD}{\fonticon{laptop-code}{MOD}}
\newcommand*{\NEW}{\fonticon{lightbulb}{NEW}}
\newcommand*{\RC}{\fonticon{graduation-cap}{RC}}
\newcommand*{\SRU}{\fonticon{recycle}{SRU}}
\newcommand*{\SP}{\fonticon{newspaper}{SP}}
\newcommand*{\DOMREP}{\fonticon{folder}{DOMREP}}
\newcommand*{\TEAM}{\fonticon{user-group}{TEAM}}
\newcommand*{\TEACH}{\fonticon{chalkboard-user}{TEACH}}
\newcommand*{\PM}{\fonticon{clipboard-list}{PM}}
\newcommand*{\USERS}{\fonticon{comments}{USERS}}
\usepackage[acronym,toc,shortcuts,nogroupskip]{glossaries}
\newglossary[skills.glg]{skills}{skills.gls}{skills.glo}{Skill codes}
\makeglossaries
\newglossaryentry{Fortran}{name={Fortran},description={general-purpose compiled programming language}}
\newglossaryentry{Cpp}{name={C++},description={general-purpose compiled programming language}}
\newglossaryentry{C}{name={C},description={general-purpose compiled programming language}}
\newglossaryentry{R}{name={R},description={general-purpose scripting language}}
\newglossaryentry{Python}{name={Python},description={general-purpose scripting language}}

\newglossaryentry{GitHub}{name={GitHub},description={online software repository hosting and collaboration platform}}
\newglossaryentry{GitLab}{name={GitLab},description={online software repository hosting and collaboration platform}}

\newglossaryentry{design-pattern}{name={design pattern},description={general and reusable solution to solve a \ac{SE} problem (often a best practice, or a ``recipe'')}}
\newglossaryentry{static-analysis}{name={static analysis},description={automated procedure to detect software bugs in source code without executing the code}}
\newglossaryentry{bus-factor}{name={bus factor},description={vulnerability of a project to losing key and irreplaceable team members - a bus factor of 1 means that a single such team member vanishing would already stall the project's advancement.}}

\newglossaryentry{Forge}{name={Forge},description={A platform that integrates software repositories and other communication or project management tools}}
\newglossaryentry{SysOp}{name={SysOp},description={system administrator in charge of a computing infrastructure}}
\newglossaryentry{DevOps}{name={DevOps},description={set of practices and tools for code development and computing infrastructure maintenance}}
\newglossaryentry{software-publication}{name={software publication},description={the practice of long-term archiving software artifacts with software metadata under a permanent identifier}}

\newacronym{EMBL-EBI}{EMBL-EBI}{European Molecular Biology Laboratory - European Bioinformatics Institute}
\newacronym{EuroHPC-JU}{EuroHPC JU}{European High-Performance Computing Joint Undertaking}
\newacronym{UNESCO}{UNESCO}{United Nations Educational, Scientific and Cultural Organization}
\newacronym{MIT}{MIT}{Massachusetts Institute of Technology}
\newacronym{ISCB}{ISCB}{International Society for Computational Biology}
\newacronym{ENCCS}{ENCCS}{EuroCC National Competence Center Sweden}
\newacronym{PRACE}{PRACE}{Partnership for Advanced Computing in Europe}
\newacronym[first={German Research Foundation (DFG, Deutsche Forschungsgemeinschaft)}]{DFG}{DFG}{German Research Foundation (Deutsche Forschungsgemeinschaft)}
\newacronym[first={National Research Data Infrastructure (NFDI, Nationale Forschungsdateninfrastruktur)}]{NFDI}{NFDI}{National Research Data Infrastructure (Nationale Forschungsdateninfrastruktur)}
\newacronym{UNIVERSE-HPC}{UNIVERSE-HPC}{Understanding and Nurturing an Integrated Vision for Education in RSE and \glsentrytext{HPC}}

\newacronym[plural=IDEs,firstplural=integrated development environments (IDE)]{IDE}{IDE}{integrated development environment}
\newacronym[plural=KPIs,firstplural=key performance indicators (KPIs)]{KPI}{KPI}{key performance indicator}
\newacronym[plural=LLMs,firstplural={large language models (LLMs)}]{LLM}{LLM}{large language model}
\newacronym{UML}{UML}{Unified Modelling Language}
\newacronym{HPC}{HPC}{High-Performance Computing}
\newacronym{HTC}{HTC}{High-Throughput Computing}
\newacronym{STEM}{STEM}{science, technology, engineering and mathematics}
\newacronym{SE}{SE}{software engineering}
\newacronym{FAIR}{FAIR}{Findability, Accessibility, Interoperability and Reusability}
\newacronym{IT}{IT}{information technology}
\newacronym{ML}{ML}{machine learning}
\newacronym{GPL}{GPL}{GNU General Public License}
\newacronym{CI}{CI}{continuous integration}
\newacronym{CD}{CD}{continuous delivery}
\newacronym{CICD}{CI/CD}{continuous integration and continuous delivery}
\newacronym{RDM}{RDM}{research data management}
\newacronym{GDPR}{GDPR}{General Data Protection Regulation}
\newacronym{PI}{PI}{Principal Investigator}
\newacronym{TDD}{TDD}{test-driven development}
\newacronym{HIFIS}{HIFIS}{Helmholtz Federated IT Services}
\newacronym{DMP}{DMP}{data management plan}
\newacronym{EU}{EU}{European Union}
\newacronym{UK}{UK}{United Kingdom}
\newacronym{GREENER}{GREENER}{Governance, Responsibility, Estimation, Energy and embodied impacts, New collaborations, Education and Research}

\newglossaryentry{DOCBB}{name={\DOCBB},type={skills},description={Creating documented code building blocks}}
\newglossaryentry{DIST}{name={\DIST},type={skills},description={Building distributable software}}
\newglossaryentry{SWLC}{name={\SWLC},type={skills},description={Adapting to the software life cycle}}
\newglossaryentry{SWREPOS}{name={\SWREPOS},type={skills},description={Use software repositories}}
\newglossaryentry{MOD}{name={\MOD},type={skills},description={Software behaviour awareness and analysis}}
\newglossaryentry{NEW}{name={\NEW},type={skills},description={Conducting and leading research}}
\newglossaryentry{RC}{name={\RC},type={skills},description={Understanding the research cycle}}
\newglossaryentry{SRU}{name={\SRU},type={skills},description={Software re-use}}
\newglossaryentry{SP}{name={\SP},type={skills},description={Software publication and citation}}
\newglossaryentry{DOMREP}{name={\DOMREP},type={skills},description={Using domain repositories/directories}}
\newglossaryentry{TEAM}{name={\TEAM},type={skills},description={Working in a team}}
\newglossaryentry{TEACH}{name={\TEACH},type={skills},description={Teaching}}
\newglossaryentry{PM}{name={\PM},type={skills},description={Project management}}
\newglossaryentry{USERS}{name={\USERS},type={skills},description={Interaction with users and other stakeholders}}
\author[,1]{Florian Goth\texorpdfstring{\thinspace\orcidlink{0000-0003-2707-4790}}{}\texorpdfstring{\thanks{Corresponding author}}{}}
\author[2]{Renato Alves\texorpdfstring{\thinspace\orcidlink{0000-0002-7212-0234}}{}}
\author[3]{Matthias Braun\texorpdfstring{\thinspace\orcidlink{0000-0001-8591-3690}}{}}
\author[4]{Leyla Jael Castro\texorpdfstring{\thinspace\orcidlink{0000-0003-3986-0510}}{}}
\author[5,6]{Gerasimos Chourdakis\texorpdfstring{\thinspace\orcidlink{0000-0002-3977-1385}}{}}
\author[7]{Simon Christ\texorpdfstring{\thinspace\orcidlink{0000-0002-5866-1472}}{}}
\author[8]{Jeremy Cohen\texorpdfstring{\thinspace\orcidlink{0000-0003-4312-2537}}{}}
\author[9]{Stephan Druskat\texorpdfstring{\thinspace\orcidlink{0000-0003-4925-7248}}{}}
\author[10]{Fredo Erxleben\texorpdfstring{\thinspace\orcidlink{0000-0001-7233-3005}}{}}
\author[11]{Jean-Noël Grad\texorpdfstring{\thinspace\orcidlink{0000-0002-5821-4912}}{}}
\author[12]{Magnus Hagdorn\texorpdfstring{\thinspace\orcidlink{0000-0002-5076-4864}}{}}
\author[13]{Toby Hodges\texorpdfstring{\thinspace\orcidlink{0000-0003-1766-456X}}{}}
\author[10]{Guido Juckeland\texorpdfstring{\thinspace\orcidlink{0000-0002-9935-4428}}{}}
\author[14]{Dominic Kempf\texorpdfstring{\thinspace\orcidlink{0000-0002-6140-2332}}{}}
\author[15]{Anna-Lena Lamprecht\texorpdfstring{\thinspace\orcidlink{0000-0003-1953-5606}}{}}
\author[16]{Jan Linxweiler\texorpdfstring{\thinspace\orcidlink{0000-0002-2755-5087}}{}}
\author[17]{Frank Löffler\texorpdfstring{\thinspace\orcidlink{0000-0001-6643-6323}}{}}
\author[18]{Michele Martone\texorpdfstring{\thinspace\orcidlink{0000-0003-3239-8554}}{}}
\author[19]{Moritz Schwarzmeier\texorpdfstring{\thinspace\orcidlink{0000-0001-8992-6245}}{}}
\author[20]{Heidi Seibold\texorpdfstring{\thinspace\orcidlink{0000-0002-8960-9642}}{}}
\author[21,22]{Jan Philipp Thiele\texorpdfstring{\thinspace\orcidlink{0000-0002-8901-6660}}{}}
\author[23]{Harald von Waldow\texorpdfstring{\thinspace\orcidlink{0000-0003-4800-2833}}{}}
\author[24]{Samantha Wittke\texorpdfstring{\thinspace\orcidlink{0000-0002-9625-7235}}{}}
\affil[1]{Würzburg-Dresden Cluster of Excellence ct.qmat, University of Würzburg, 97074, Würzburg, Germany}
\affil[2]{European Molecular Biology Laboratory, Heidelberg, Germany}
\affil[3]{Cluster of Excellence IntCDC, University of Stuttgart, Germany}
\affil[4]{ZB MED Information Centre for Life Sciences, Cologne, Germany}
\affil[5]{School of Computation, Information and Technology, Technical University of Munich, Garching, Germany}
\affil[6]{Institute for Parallel and Distributed Systems, University of Stuttgart, Stuttgart, Germany}
\affil[7]{Leibniz University Hannover, Department of Cell Biology and Biophysics, Computational Biology, Germany}
\affil[8]{Imperial College London, London, UK}
\affil[9]{German Aerospace Center (DLR), Institute for Software Technology, Berlin, Germany}
\affil[10]{Helmholtz-Zentrum Dresden-Rossendorf, Germany}
\affil[11]{Institute for Computational Physics, University of Stuttgart, Germany}
\affil[12]{Geschäftsbereich IT, Charité Universitätsmedizin Berlin, Germany}
\affil[13]{The Carpentries, USA}
\affil[14]{Heidelberg University, Scientific Software Center, Germany}
\affil[15]{Institute of Computer Science, University of Potsdam, Germany}
\affil[16]{Technische Universität Braunschweig, Germany}
\affil[17]{Michael Stifel Center Jena \& Friedrich Schiller University Jena, Germany}
\affil[18]{Leibniz Supercomputing Centre, Garching, Germany}
\affil[19]{Mathematical Modeling and Analysis, TU Darmstadt, Germany}
\affil[20]{IGDORE Munich, Germany}
\affil[21]{Weierstrass Institute, Berlin, Germany}
\affil[22]{Leibniz University Hannover, Institute of Applied Mathematics, Scientific Computing, Hannover, Germany}
\affil[23]{Johann Heinrich von Thünen Institute, Centre for Information Management, Germany}
\affil[24]{CSC – IT Center for Science, Espoo, Finland}

\usepackage[capitalise]{cleveref}
\ifLuaTeX
  \usepackage{selnolig}  % disable illegal ligatures
\fi
\usepackage[alldates=iso]{biblatex}
\usepackage{bookmark}
\IfFileExists{xurl.sty}{\usepackage{xurl}}{} % add URL line breaks if available
\hypersetup{
  pdftitle={Foundational Competencies and Responsibilities of a Research Software Engineer},
  pdfkeywords={research software
engineering, RSE, competencies, curriculum design, teaching},
  colorlinks=true,
  linkcolor={darkgray},
  filecolor={Maroon},
  citecolor={Blue},
  urlcolor={Blue},
  pdfcreator={LaTeX via pandoc}}

\title{Foundational Competencies and Responsibilities of a Research
Software Engineer}
\usepackage{etoolbox}
\makeatletter
\providecommand{\subtitle}[1]{% add subtitle to \maketitle
  \apptocmd{\@title}{\par {\large #1 \par}}{}{}
}
\makeatother
\subtitle{Current State and Suggestions for Future Directions}
\date{2025-07-15}

\begin{document}
\maketitle
\vspace{-1em}
\begin{abstract}
The term Research Software Engineer, or RSE, emerged a little over 10
years ago as a way to represent individuals working in the research
community but focusing on software development. The term has been widely
adopted and there are a number of high-level definitions of what an RSE
is. However, the roles of RSEs vary depending on the institutional
context they work in. At one end of the spectrum, RSE roles may look
similar to a traditional research role. At the other extreme, they
resemble that of a software engineer in industry. Most RSE roles inhabit
the space between these two extremes. Therefore, providing a
straightforward, comprehensive definition of what an RSE does and what
experience, skills and competencies are required to become one is
challenging. In this community paper we define the broad notion of what
an RSE is, explore the different types of work they undertake, and
define a list of foundational competencies as well as values that
outline the general profile of an RSE. These foundational skills are
encountered to a large extent within the skill sets of current RSEs in
Germany and beyond, and we propose them as a starting point for aspiring
RSEs to shape their technical profile. Further research and training can
build upon this foundation of skills and focus on various aspects in
greater detail. We expect that graduates and practitioners will have a
larger and more diverse set of skills than outlined here. On this basis,
we elaborate on the progression of these skills along different
dimensions. We look at specific types of RSE roles, propose
recommendations for organisations, and give examples of future
specialisations. An appendix details how existing curricula fit into
this framework.
\end{abstract}

\newpage

\newgeometry{top=2.5cm,right=2.5cm,bottom=2.5cm,left=2.5cm}

{
\hypersetup{linkcolor=}
\setcounter{tocdepth}{3}
\tableofcontents
}
\section{Introduction}\label{sec:introduction}

Computers and software have played a key role in the research life cycle
for many decades. They are now vital elements of the research process
across almost all domains. They enable researchers to collect and
process ever-increasing amounts of data, simulate a wide range of
physical phenomena across previously unexplored scales of the universe,
and discover previously inconceivably complex structures in nature and
societies via \ac{ML}. This prevalence of computation and
digitally-aided data analysis in research means that digital skills are
now required by researchers at all career levels, and in fields
significantly beyond those that would previously have been expected.
Research software is now used and developed not only in \ac{STEM}
domains, but also in other fields, like medicine and the humanities.

Researchers often lack the skills to use specialised software for their
research, let alone write it~\autocite{NamingPain}. If they come from a
non-technical domain, they may also struggle to know what to ask when
trying to request help from and interact with more experienced
colleagues. A gap still exists in academic education, as many curricula
do not sufficiently prepare students in this regard. This situation is
exemplified by the extracurricular \ac{MIT} class ``The Missing Semester
of Your CS Education''~\autocite{MIT}, which aims to increase
``computing ecosystem literacy'' even among students of Computer Science
at \ac{MIT}.

Researchers investing increasing amounts of their time developing their
\ac{SE} skills to support their research work can find themselves with
little time to do the research itself. This, in turn, presents career
development challenges since the experience required to gain and
progress in research and academic roles is traditionally assessed
through metrics that do not directly include software outputs. A recent
shift towards the establishment of the distinct role of a
\emph{``Research Software Engineer''}~\autocite{WhatResearchSoftware}
(RSE, a term that was coined in the \ac{UK} a little over 10 years
ago~\autocite{Hettrick2016}), now provides a base on which sustainable
career opportunities can be (and are being) built, allowing for better
training of researchers and more effective support for the development
of high quality research software. There is still a long way to go, but
positive change is well underway.

RSEs may work within one of the increasing number of research software
engineering teams that have been set up at universities and research
organisations over the past decade, or they may be embedded within a
research team. They may have a job title that officially recognises them
as an RSE, or they may have a standard research or technical job title
such as Research Assistant, Research Fellow, or Software Engineer.
Regardless of their job title, RSEs share a set of core skills that are
required to design and develop research software, understand the
research environment, and ensure that they produce sustainable,
maintainable code that supports reproducible research outputs, following
the \ac{FAIR} principles~\autocite{FAIR4RS}.

This community paper defines a set of core values and foundational
competencies, which an RSE should acquire during training and formal
education. These skills are formulated independently of a specific
research domain and current technical tools used to support the
application of the skills. By defining these competencies, we provide a
guiding framework to facilitate the training and continuous professional
development of RSEs, thus helping to provide a positive impact on
research outputs and, ultimately, society as a whole. These competencies
draw upon skills from traditional SE practice, established research
culture, and the commitment to being part of a team. However, we see
this set of skills as a foundation to build upon. We envision that
through specialised training, the set of skills of graduate RSEs and
domain researchers will grow. This is underlined by a growing interest
to perform RSE research, i.e.~research into methods and tools more
catered to the unique challenges that research software provides.

While this community paper is based on workshop discussions that were
attended largely by RSEs (deRSE23 in Paderborn \autocite{derse23pad},
un-deRSE23 in Jena, and deRSE24 in Würzburg, all in Germany), we believe
that the competencies formulated here can offer far-reaching impact
beyond the domain of RSE into adjacent aspects of research and, indeed,
the wider research community. This is especially important given that
much research involves some amount of data management, processing and
visualisation, or the creation of tools for these tasks, and funding
bodies and computing infrastructure providers will sometimes prioritise
projects that generate archived, annotated, re-usable, and potentially
remotely executable data. In particular, funding agencies and research
managers will find the discussion in this paper valuable in order to
discover where RSEs see their place in the existing landscape of
scientific domains and how to support the work of RSEs at different
positions and career levels. While we draw mostly from experiences of
RSEs working in Germany or, in some cases, across Europe and the US, our
recommendations do not focus on a particular region.

The outline of the paper is as follows. We start with a non-exhaustive
overview of existing initiatives in \cref{sec:related-work}.
\cref{sec:values} elaborates on the values that provide the guiding
principles for the work of an RSE. \cref{sec:required-generic-skills}
defines a set of core skills based on these values. We categorise these
skills into three pillars, namely ``software/technical'', ``research'',
and ``communication'' skills, reflecting the hybrid nature of an RSE. To
justify the selection of these skills, we also list some current tasks
and discuss the skills used therein. As with any general skill set, not
all RSEs will need to use all the skills highlighted to the same level
of expertise. Therefore, \cref{sec:how-much-to-know} examines how much a
person needs to know depending on their education or career level or on
the type of projects they would like to be involved with. In the same
section, we provide an overview of what skills and limitations an RSE in
different team structures typically has, and we give recommendations for
organisations that need to support RSEs. \cref{sec:rse-specialisations}
provides a list of RSE specialisations and discusses the level of skill
needed to work in each of them, before we conclude the paper with
details of future work in \cref{sec:future-work} and conclusions in
\cref{sec:conclusion}. Finally, \cref{sec:appendix} provides an example
curriculum in \cref{subsec:examplemaster}, a story-like description of a
fictional RSE career progression in \cref{subsec:examplecareer}, and a
list of existing skills and certifications in related fields, in
\cref{subsec:existingframeworks}.

\subsection{Terminology}\label{terminology}

\subsubsection{The term Research Software
Engineer}\label{the-term-research-software-engineer}

Research Software Engineering can be considered an interface discipline,
linking traditional Software Engineering with Research
itself~\autocite{Lamprecht2024-giradar}. Due to this nature there is a
plethora of different variations of RSE depending on the particular
Research domain they are working in. Therefore the broad notion of
Research Software Engineers is better thought of as a collection of
sub-communities. The term Research Software Engineer is made more
difficult to grasp since an internationally recognised definition is
still missing. While there is consensus about the general notion that an
RSE is a person with one leg in their research domain and the other in
software development, this spans a whole spectrum depending on which one
is more emphasised. There is also the question of what level of
professionalism concerning both non-SE research and SE is expected. A
more inclusive definition allows more people to self-identify as RSEs,
thereby also fostering an inclusive community of people working in
digital science (see also~\cref{sec:values} on the values of an RSE).
RSEs fall therefore somewhere on the spectrum between a researcher at
one end and a software engineer at the other. Common to all of them is,
that they need to be able to work in the research environment the
software is used in, ideally at eye-level with native researchers, but
at least as close as possible. RSEs often need to deal with
non-technical complexities that are characteristic for research
environments: organisational, motivational, with respect to the size of
projects, independence and heterogeneous goals of stakeholders, boundary
conditions for funding and future funding, to name just a few.
Summarising, RSEs have skills and experience in three important areas:
in the research area(s) their software is used in, in software
engineering topics, as well as in interdisciplinary communication.

\subsubsection{Further definitions}\label{further-definitions}

Depending on the national research environments and processes that
readers are familiar with, the notion of the terms \emph{software} and
\emph{research} might differ. Therefore, to avoid ambiguities, we define
these as follows:

\textbf{Software}: Source code, documentation, tests, executables and
all other artefacts that are created during the development process that
are necessary to understand its purpose.

\textbf{Research software}: Foundational algorithms, the software
itself, as well as scripts and computational workflows that were created
during the research process or for a research purpose, across all
domains of research. This definition is broader than
in~\autocite{FAIR4RS} and is the outcome of a recent discussion
in~\autocite{Gruenpeter2021}.

\textbf{Research software engineers}: People who create or improve
research software and/or the structures that the software interacts with
in the computational environment of a research domain. They are highly
skilled team members who may also choose to conduct their own research
as part of their role. However, we also recognise that many RSEs have
chosen specifically to focus on a technical role as an alternative to a
traditional research role because they enjoy and wish to focus on the
development of research software.

\textbf{Researchers}: People who are using the services provided by
Research Software Engineers. This, on purpose, is a very broad
definition and was chosen for a better reading.

\section{Related work}\label{sec:related-work}

Various initiatives are working to support technical professionals
develop their computational skills. Particularly related to this work
are initiatives that aim to define sets of such skills and to guide the
community with certification programs and training resources.

\paragraph{RSE Competencies Toolkit}\label{rse-competencies-toolkit}

The RSE Competencies Toolkit~\autocite{RSECompetenciesToolkit2023} is a
community project that developed out of a hack day activity at the 2023
edition of the annual Software Sustainability Institute Collaborations
Workshop~\autocite{SSICW23}. The toolkit provides a web application that
aims to support technical professionals in understanding how to develop
their skills. It enables them to build a profile of their competencies
within the system, while it also provides a set of training resources
that are linked to a competency framework.

\paragraph{HPC Certification Forum}\label{hpc-certification-forum}

The \ac{HPC} Certification Forum~\autocite{HPCCF} is working towards
providing a certification process for \ac{HPC} skills. As part of this
process, the group is developing a Competence
Standard~\autocite{HPCCFCompetencies} and an associated skill tree that
provides a classification of \ac{HPC} competencies. This work aims to
develop a standardised representation of relevant \ac{HPC} knowledge and
skills which can, in turn, lead to structured and recognised sets of
skills that can underpin the certification process.

\paragraph{EMBL-EBI Competency Hub}\label{embl-ebi-competency-hub}

The \ac{EMBL-EBI} Competency Hub~\autocite{CompetencyHub} provides a
bioinformatics/computational biology-focused example of a competency
portal. In addition to collecting information on a range of competencies
that can be browsed within the web-based tool, it also provides career
profiles for roles within the domains that \ac{EMBL-EBI} focuses on. The
hub provides access to a variety of training resources that are linked
to the specific competencies that they relate to. This enables learners
to more easily find the right training materials in order to support
their career development journey, helping them to identify what they
might want to learn and in what order.

\paragraph{Training-focused
initiatives}\label{training-focused-initiatives}

Further initiatives implicitly define sets of competencies by providing
(open) teaching material for selected skills. This is a non-exhaustive
list of related initiatives, which will be discussed in more detail in a
separate publication. In some cases, the activities extend beyond
training, but they do not focus on defining frameworks of competencies.

One prominent example is the Carpentries~\autocite{Carpentries}, a
non-profit entity that supports a range of open source training
materials and international communities of volunteer instructors and
helpers who run courses around these materials. A similar framework is
provided by CodeRefinery~\autocite{CodeRefinery}, currently funded by
the Nordic e-Infrastructure, as well as
SURESOFT~\autocites{SURESOFTLink}[~][]{SURESOFT2022}, a project at
Technical University (TU) Braunschweig and
Friedrich-Alexander-University (FAU) Erlangen-Nürnberg, funded by the
\ac{DFG} and targeting more advanced \ac{SE} topics such as software
design principles, \glspl{design-pattern}, refactoring, \ac{CI} and
\ac{TDD}. The INTERSECT RSE Training project
\autocite{INTERSECTOnlineResources,Carver2020} also provides training
materials and organises training events in the USA, funded by the NSF.

There are also several initiatives focused on training
\acrshort{HPC}-oriented RSEs, such as the
\acrfull{PRACE}~\autocite{PRACE} (with material aggregated on various
websites, e.g., on EuroCC Training~\autocite{EuroCCTraining}),
\ac{UNIVERSE-HPC}~\autocite{UNIVERSEHPC} (a project funded under the
\acrshort{UK}'s ExCALIBUR research programme~\autocite{EXCALIBUR}), and
the \ac{ENCCS}~\autocite{ENCCS}, which offers a collection of lessons
for \ac{HPC} skills~\autocite{ENCCSLessons}. At the intersection between
\ac{HPC} and the broader RSE field, the IDEAS PRODUCTIVITY project
\autocite{IDEAS} organises online events, provides training material via
the Better Scientific Software (BSSw) project \autocite{BSSW} and
maintains HPC-focused guidelines, such as the Extreme-scale Scientific
Software Development Kit \autocite{xSDK}.

Initiatives focused on Germany include EduTrain~\autocite{EDUTRAIN} (a
section of the \acrfull{NFDI}~\autocite{NFDI}), the
\acrfull{HIFIS}~\autocite{HIFIS}, and the already mentioned
SURESOFT~\autocite{SURESOFTLink}.

\section{Values}\label{sec:values}

It is important that the activities of an RSE are guided by ethical
values. In addition to the values for good scientific practice
\autocite{dfg_gsp}, RSEs also need to adhere to the \ac{SE} Code of
Ethics \autocite{Gotterbarn1999}. Central to that code is the RSE's
obligation to commit to the health, safety and welfare of the public and
act in the interest of society, their employer and their clients.
Further values loosely based on that code include the obligations

\begin{itemize}
\tightlist
\item
  to commit to objectivity and fact-based, honest research conclusions,
\item
  to promote openness and accountability in the research process,
\item
  to take great care to develop software that adheres to current best
  practices,
\item
  to judge independently and maintain professional integrity,
\item
  to treat colleagues and collaborators with respect and work towards a
  fair and inclusive environment, and
\item
  to promote these values whenever possible and make sure that they are
  passed on to new practitioners.
\end{itemize}

Many practitioners will follow the values expressed in these codes of
conduct without knowing them because they are passed on implicitly by
their peers and mentors \autocite{Consoli2008}. Here, they are stated
explicitly because they underpin the foundational competencies and
responsibilities of RSEs who are professionals living in both worlds.

The deployment of computer-based modelling and simulation has
dramatically changed the practice of science in a large number of
fields. It has enabled the hitherto impossible study of new classes of
problems, often replacing traditional experimentation and observation
(it can also serve to integrate a communal body of
knowledge~\autocite{Parker2022}). Thereby it has the potential of
changing our way of generating knowledge, while at the same time it
challenges our notions of explaining science.
Humphreys~\autocite{humphreys_extending_2004} regards this development
as ``more important than the invention of calculus in the 1660s, an
event that remained unparalleled for almost 300 years''. The
epistemological status of computer modelling and simulation is still the
subject of debate, which ranges from the postulate of a new process of
knowledge creation that has its own, unique,
epistemology~\autocite{winsberg_sanctioning_1999} to the perception that
from a philosophy of science perspective, there is nothing really
new~\autocite{frigg_philosophy_2009}. In any case, it is clear that a
number of decisions in the construction of a simulation-model will have
a significant impact on the adequacy for
purpose~\autocite{bokulich_data_2021} of the model. These decisions
include the selection of the salient characteristics of the system to be
modelled, the choice of the mathematical representation of the processes
to be represented, the choice of numerical methods and other algorithms
and even including the design of the user-interface.

The relationship between initial state, inputs and final state of a
computer simulation is ``epistemically
opaque''~\autocite{humphreys_extending_2004}, in that not every step of
the process is directly observable. The current trend of an increasing
application of computationally irreducible systems, such as those based
on artificial neural networks, further exacerbates this inherent
limitation of explainability. An RSE usually takes a pivotal role in
assessing this adequacy for purpose of a model as well as in
characterising and communicating the domain of its legitimate
application and its limits of interpretability. This role, together with
the enormous reliance on modelling and simulation of scientific results,
as well as real-world decision-making, places a large responsibility on
the RSE. It is important that RSEs are aware of this responsibility and
continuously improve their capabilities to live up to it.

Research software is also well on its way to being ever-present in
data-driven research, in all research fields. This can probably be most
prominently seen by considering software used to analyse data,
e.g.~within experimental research. It is not unusual for RSEs to support
those more research data oriented efforts as well. Here, specifically,
they closely interact with research data management professionals and
practices by designing research software that is better able to adhere
to the \ac{FAIR} principles for research data, but also to follow
similar rules for research software (FAIR4RS \autocite{FAIR4RS}). As
such, they are then familiar with special requirements stemming from the
field itself, e.g., in medical research, and with privacy related issues
especially for personal data, e.g., for conducting surveys.

RSEs often assume a multifaceted role at the junction of research,
\ac{SE} and data management. They work with a varying and diverse set of
colleagues that might include other developers, support unit staff and
academics of different fields and all career stages. This situation
yields a specific set of challenges RSEs should be aware of to
consciously make ethically sound judgement calls. Below we list some
example areas that highlight present-day challenges.

\subsection{Current challenges}\label{current-challenges}

\subsubsection{Data security}\label{sec:personal-data}

A lot of RSE work involves the manipulation or creation of data
processing tools. We highlight that professional conduct requires these
creations to be reliable and to maintain data integrity. In particular,
the way that personal data is handled can have far-reaching implications
for society. Independent of the encoding into the respective national
law in an RSE's jurisdiction, the right to information privacy is
internationally recognised as a fundamental human right, e.g., in the
European Convention on Human
Rights~\autocites{CouncilOfEurope-ETS005-2021}[~][]{Hirvela2022}. RSEs
need to be aware of this topic's importance and deal with tensions that
might arise with researchers' desire for trouble-free sharing of data,
thereby expecting openness about the research process, versus the
integrity expectations of the society towards \ac{IT} systems. Handling
personal data also has ramifications for information security
considerations during the software development process. Data protection
is a complex topic, so RSEs should be aware that they may need to
consult external expertise, for example when dealing with special topics
such as cryptography or re-identification
attacks~\autocite{Henriksen2016}.

\subsubsection{Mentoring and
diversity}\label{sec:mentoring-and-diversity}

RSEs are often experienced professionals who work closely with and
provide technical training and guidance to early career researchers.
Similarly to academic supervisors, they bear a certain responsibility to
guide and advise less-experienced colleagues with respect to career
development and the achievement of academic goals. This can take the
form of supervising a student or mentoring a fellow RSE. The RSE needs
to be aware of the biases arising from the sociological imbalances in
research and academia. According to the \ac{UNESCO} Science Report
\autocite{Schneegans2021} women account for 33.3\% of all researchers.
60.2\% of researchers come from high-income countries which account for
17.5\% of the global population in 2018. Furthermore, the socioeconomic
background of academics is not representative of the general population,
for example in the US a tenure-track academic is 25 times more likely to
have a parent with a PhD \autocite{Morgan2022}. Thereby, to promote
their values of an honest, open, and inclusive research space, they
should be aware of the diversity problems and help to mitigate them
whenever they have the chance to do so.

\subsubsection{Shaping digital science}\label{shaping-digital-science}

Through writing research software, RSEs hold an important role in the
process of scientific production. Their choices might determine whether
the respective research is reproducible or not, whether the results can
be re-used, whether future research can build on existing tools or has
to start from scratch. Builders of larger research-infrastructure
projects determine to some extent the possibilities and limitations of
future research and therefore need to be able to make a value-based
judgement on topics such as open science, path dependence, and vendor
lock-in.

\subsubsection{Addressing environmental sustainability within planetary
limits}\label{sec:environmental-sustainability}

The last two decades saw transistor technology approach the limits of
attainable miniaturisation, and maximum chip clock frequency begin to
plateau~\autocite{Sutter2005}. Nevertheless, a misleading belief in
limitless growth of computing capabilities (storage, computing power,
transfer speed) is still widespread within popular perception. A
practical consequence of this is an ever-growing demand for resources to
cover the expanding need of storage and processing, with no clear
deceleration in sight (e.g.~the IEA estimates a doubling in data centres
energy consumption from 2024 to 2026~\autocite{IEA2024}). At the same
time, current science is well aware of several planetary boundaries
being exceeded due to human activities \autocite{Richardson2023}. Data
processing, storage and transfer account for a non-negligible
fraction~\autocite{IEA2024}. Demands to move resource consumption to a
sustainable rate are well justified and supported by
science~\autocite{Sills2019}.

RSEs have the opportunity to contribute to this effort by, for example,
choosing computationally adequate approaches (e.g.~recognising where a
proven statistical method may suffice in place of a power-hungry AI
model, or configuring a test pipeline to minimise redundancy), and
embracing data frugality measures (e.g.~recognising sufficient
resolution when sampling data for processing or storage). If past
computational solutions were frugal because of technological limits, in
future they should tend to that by virtue of an awareness of what may be
adequate. The \ac{GREENER} principles~\autocite{Lannelongue2023} suggest
how these concerns can be addressed and how research computing can
become more environmentally sustainable.

\subsubsection{Emerging challenges}\label{emerging-challenges}

RSEs often operate at the cutting edge of technological development and
therefore might have to deal with technologies of which the dangers and
drawbacks are still poorly understood. A current example is the rush for
the application of \acp{LLM}, where RSEs working in these fields should
stay up-to-date and be able to help researchers assess topics such as
training-data bias, \ac{LLM} ``hallucinations'' or malicious use, with
the greater goal of making these powerful tools work for the welfare of
society.

\section{Foundational RSE
competencies}\label{sec:required-generic-skills}

The role of an RSE lies somewhere on the spectrum between that of a
researcher (the ``R'') and a software engineer (the ``SE'') and,
therefore, requires competencies in both fields. RSEs typically have a
background in research or software engineering, but they definitely have
obtained broader knowledge in both fields. Even when working as the only
RSE on a task or project, they typically apply their knowledge and
experience as part of larger teams of researchers and technical
professionals, which allows them to cultivate this hybrid nature. There
are many ways to categorise the competencies of an RSE. We chose to
distribute these competencies over three pillars to reflect the fact
that RSEs are both competent researchers (the research skills,
\cref{sec:research-skills}) and software engineers (the
software/technical skills, \cref{sec:software-skills}). The third pillar
(communication skills, \cref{sec:communication-skills}) forms the bridge
between the former two categories, with a particular focus on the
software and research cycle and the scientific process. These
competencies are relevant in a broad setting and form the foundation for
specific specialisations. These competencies have been chosen in order
to make RSEs contribute to an open and inclusive research environment,
with tools that respect their professional values (see
\cref{sec:values}).

These skills and competencies come into play in various forms: The RSEs
themselves need to acquire and develop them as their career progresses
(\textbf{Career level}). However, some knowledge of software and data
processing is required at all academic levels and for all positions
(\textbf{Academic Progression}). The relative importance of the skills
and competencies also depends on the size of the RSE team
(\textbf{Project team size}). Finally, different sets of skills are
emphasised in the different RSE specialisations (\textbf{RSE
specialisations}).

During the Paderborn workshop (deRSE23) we asked learners and novice
RSEs what they would like to have learnt. The top five items mentioned
were (\textcite{derse23pad}): testing, contributing to large projects,
when or why to keep repositories private, high-quality software
development, and finding a community. Those topics comprise combinations
of the skills and competencies defined below. We will elaborate these in
\cref{sec:tasks-and-responsibilities}.

\subsection{Software/Technical skills}\label{sec:software-skills}

\newcommand{\skillsection}[1]{\hypertarget{skills-#1}{%
\subsubsection{\glsentrydesc{#1} (\texorpdfstring{\glsentrytext{#1}}{#1})}\label{skills-#1}}}

Besides skilled researchers, RSEs are also competent software engineers.
As such, they ideally can solve complex software engineering problems
and design software as a user-oriented, future-proof product. The
technical skills required by an RSE overlap to a large extent with the
common fundamental software engineering skills (see, e.g.,
\textcite{Landwehr2017}), but put greater emphasis on aspects related to
achieving good scientific practice and to serving special needs of
research software. In addition, a lot of RSEs are either self- or peer
taught in these skills (see, e.g., figure~14 in~\textcite{Barker2023}).
These skills include requirements analysis, design, construction,
testing, program analysis, and maintenance of software. On the other
hand, RSEs also know how to make research software adhere to the
\ac{FAIR} principles \autocite{FAIR4RS}, and how to achieve different
levels of research software reusability (see, e.g.,
\textcite{ChueHong2014}), while they have deeper understanding of the
scientific context around the research software projects they work on.
To reflect this, the technical skills listed below complement
competencies regarding the standard life cycle of software development
(as summarised in \autoref{subsec:technical-general}) with RSE-specific
focus skills.

\subsubsection{Classical software engineering skills}\label{subsec:technical-general}

To summarise the vast range of the skills a software engineer is
typically equipped with, we refer to the Guide to the Software
Engineering Body of Knowledge (\textcite{swebok_2014}). Because research
software engineering is an interface discipline, RSEs are often stronger
in topics more commonly encountered in research software contexts (e.g.,
mathematical and engineering foundations) than in other areas (e.g.,
software engineering economics). However, they bring a solid level of
competence in all software engineering topics. Therefore, RSEs can set
and analyse software requirements in the context of open-ended,
question-driven research. They can design software so that it can
sustainably grow, often in an environment of rapid turnover of
contributors. They are competent in implementing solutions themselves in
a wide range of technologies fit for different scientific applications.
They can formulate and implement various types of tests, they can
independently maintain software and automate operations of the
integration and release process. They can provide working, scalable, and
future-proof solutions in a professional context and with common project
and software management techniques, adapted to the needs of the research
environment. Finally, as people who have often gained significant
research experience in a particular discipline, they combine the
necessary foundations from their domain with software engineering skills
to develop complex software.

\skillsection{SWLC}

The traditional software development life cycle defines the stages that
form the process of building a piece of software. Initial development
generally involves an analytic process where requirements and ideas are
gathered and analysed (requirements engineering), followed by
formulating a plan to fulfil them (design) that is finally turned into
running code (implementation). This is accompanied by different measures
of quality control (e.g., reviews, testing), validating and verifying
that things work as expected and that they continue to do so when
development progresses further. Depending on the software project, this
can mean a simple ``Think-before-you-do'', or more elaborate and formal
processes\autocite{mundt2022tiered}. Often the development cycles are
executed iteratively and incrementally. The life cycle further includes
periods of deployment, maintenance and further development (software
evolution), as well as software retirement. To assess the current state
and needs of the software, the RSE should be familiar with different
maturity metrics, e.g.~the DLR application
classes~\autocite{Schlauch2018b}, the research software maturity
model~\autocite{Deekshitha2024} or technology readiness levels~(TRLs).
Additionally, the research software life cycle extends the traditional
life cycle with \gls{software-publication}. The RSE should be aware of
this life cycle and be able to predict and cater to the changing needs
of a software project as it moves through the stages.

\skillsection{DOCBB}

The RSE should be able to create building blocks from source code that
are reusable. This ranges from simple libraries of functions up to
complex architectures consisting of multiple software packages. An
important part of enabling code reusability is the provision of
sufficient information in the form of comments within code,
documentation or other means. This is vital to ensure that developers
and maintainers understand what a piece of software aims to do and how
to enable others to use the provided functionality. This is primarily
achieved through a ``clean'' implementation and enhanced by
documentation. Documentation ranges from commenting code blocks to using
documentation (building) tools. It should be written with consideration
for the different audiences who may need it depending on their goals and
expertise, for example by following the Diátaxis
framework~\autocite{Procida_Diataxis_documentation_framework}.

\skillsection{DIST}

The RSE should be able to distribute their code on their domain/language
specific distribution platforms. This almost always encompasses
handling/documenting dependencies with other packages/libraries. It
sometimes requires knowledge of using build or package management
systems to enable interoperability with other projects. In terms of
usability and needs of the user community the RSE should be able to
decide whether a library or a framework is the right type of program to
build and distribute.

\skillsection{SWREPOS}

The RSE should be able to identify and use fitting software
\glspl{Forge} (often just termed ``repos'') to share the artefacts they
have created and, if possible, invite the public to scrutinise them in
an open review process. These software repositories usually provide
facilities for software development, which differentiate them from the
domain repositories described later.

\skillsection{MOD}

We define this as a certain quality of analytical thinking that enables
an RSE to form a mental model of a piece of software in a specific
environment (program comprehension). Using that, an RSE should be able
to make predictions about the behaviour of a program. This is a required
skill for common tasks such as debugging, profiling, optimising,
designing good tests, or predicting user interaction. Many tools exist
to help with understanding and evaluating existing code, especially from
a structural point of view. An RSE should understand their output and
its implications. An important facet of this capability relates to
information security. RSEs need to consider the safety and integrity of
personal data and other sensitive information and make sure that they do
not negatively impact the integrity of their institution's network and
computing infrastructure (see \cref{sec:personal-data}).

\subsection{Research skills}\label{sec:research-skills}

\skillsection{NEW}

RSEs are curious and able to conduct research, both on research software
engineering, and on their research-wise ``home domain'' (see also
\cref{subsec:examplemaster}). Senior RSEs are also able to lead
research, and many RSEs have a
doctorate~\autocite{hettrick_survey_2022}. Since RSEs often operate in
different research fields, they also gain their reputation from their
effectiveness in interacting with researchers from the same or other
domains. Therefore, some curiosity together with a broad overview of the
research field is required, as this enables the RSE to learn new methods
and algorithms directly from domain peers. Similarly, a broad overview
of the field of SE research and the growing field of RSE research
enables the RSE to learn, apply, and teach new methods and tools for
improving the way they develop software. This curiosity, together with
the ability to convert it into new ideas, is also reflected when an RSE
is actively trying out new tools or discovering related literature from
adjacent domains. Lifelong learning is then no longer just a phrase but
becomes a motivation to work.

\skillsection{RC}

One of the key skills that RSEs have is their understanding of how
research works. They embrace being part of a larger community which,
despite friendly competition, shares the common goal of gaining
knowledge to disseminate it. Thereby they know that they are part of a
bigger undertaking that involves many other parties in and outside their
domain, and also that their software can be utilised at different stages
of the research cycle by different people. They may be asked to
contribute to the ethical and regulatory evaluation of a project to
ensure integrity of the research performed therein. Like other
researchers, RSEs are open to discussions and arguments beyond their own
expertise and appreciate the underlying principles of good research,
including publications, reviews and reproducibility.

\skillsection{SRU}

The re-use of existing assets such as libraries and pieces of code to
improve efficiency and quality belongs to the fundamentals of software
construction~\autocite{swebok_2014}. To discover software, RSEs rely on
domain-specific knowledge and domain repositories, as well as research
skills, discovering related software via software citations and
metadata. To evaluate whether the artefacts to be re-used suit their
needs, RSEs often need to consider the scientific context of their
origin. For example, a paper that references the code under
consideration might be crucial to validate its fitness for purpose or
lack of suitability. Code that incorporates research-domain specific
knowledge needs to be understood at a very detailed level and its re-use
documented to meet standards of good research practice. Not only the
technical compatibility needs to be understood and documented
(programming languages, system interoperability), but also the
underlying models and computational methods need to fit the purpose;
this question often requires wider research skills and deeper
understanding of the research domain at hand.

\skillsection{SP}

Another part of \ac{FAIR} software is concerned with publishing new and
derived works and making them available for re-use by the research
community and the general public, within the boundaries set by their
institutional policies. RSEs need to have a basic understanding of
common software licence types, including proprietary and open source
licences and how ``copyleft'' and ``permissive'' open source licences
differ. They should also understand compatibility between different
licences, and the ramifications for re-using and composing programs.
Beyond that, RSEs will need to properly execute the technicalities of
software publishing. These include the application of licences and
copyright statements, understanding and assigning software authorship,
crediting contributors, maintaining FAIR software metadata and
publishing software artefacts on respective publication platforms.
Finally, RSEs will need to understand the principles of software
citation~\autocite{smith_SoftwareCitationPrinciples2016}. This concerns
both the potential for reuse of their own work, which demands the
provision of complete and correct up-to-date citation metadata for their
software, as well as their own citation obligations deriving from
building on previous work in the form of dependencies.

\skillsection{DOMREP}

Almost all research software is developed within a specific scientific
domain. Some software may be able to cross boundaries, but the majority
will have a home domain, with which it needs to be able to interact. The
RSE then needs to be aware of any domain specific repositories that will
contain data sets, catalogues, and other domain specific artefacts, in
addition to software. The RSE also needs to be aware of how their
software can interact with the existing domain-specific data
repositories. Finally, they need to be able to assess and use software
repositories - domain-specific or generic - for publishing software with
the relevant metadata.

\subsection{Communication skills}\label{sec:communication-skills}

RSEs do not work in isolation. They are embedded in a research group or
work within a team of RSEs supporting particular research projects. RSEs
often need to interact with and facilitate communication among
colleagues, clients and contractors with a very broad spectrum of
background-knowledge, specialisation, expectations, and experience
whilst keeping diversity issues in mind
(\cref{sec:mentoring-and-diversity}). Communication skills are therefore
crucially important. Team skills are also mentioned in common guides for
\ac{SE} such as the software engineering body of
knowledge~\autocite{swebok_2014}. However, the interpersonal and
organisational skills and the capacity for adaption required to work in
a research setting warrants a much stronger emphasis on this field of
competence.

\skillsection{TEAM}

Being able to work, and effectively communicate in teams is essential
for RSEs. For example, RSEs need to be able to explain particular
implementation choices made and may even need to defend them. Within a
team of RSEs, code reviews improve knowledge transfer and increase team
cohesion. The team might change on a project-to-project basis and might
be comprised of colleagues with very different backgrounds including,
for example, \ac{IT} staff, domain scientists and technicians working
alongside software engineers. The shared values come into play and each
RSE needs to ensure that these values are lived by and passed on to
others. Senior RSEs may lead a team of RSEs.

\skillsection{TEACH}

RSEs have many opportunities to teach. These range from inducting new
colleagues to teaching digital skills either through short courses, for
example from The Carpentries~\autocite{Carpentries}, or entire lecture
series. RSEs may also act as mentors and consultants. Code review also
includes aspects of the teaching skill.

\skillsection{PM}

The RSE should have knowledge of project management processes. At some
institutes, project management tools and approaches differ between
individual research groups, but it is useful if an RSE understands
general structures of a \gls{PM} scheme, or can bring in new ideas for
improvement. Project management in research software engineering poses
specific challenges (see \gls{USERS}) that might require the capacity to
adapt to changing conditions and deviate from common project management
methods. Additionally, the RSE should know that SE offers various
methods and approaches specifically tailored to management of software
projects and products.

\skillsection{USERS}

Since research software is often developed as part of the research
process itself, its requirements and specifications might change with
the progression of research. Stakeholders of research software often
change across different research projects or even within the course of
one project. Roles in connection with research software are often in
flux and diffuse. For example, a single person might be user, developer
and project manager at the same time. Often this means it is necessary
for an RSE to think ``outside their comfort zone'', but at the same time
to be able to convey their knowledge and experience to experts of other
fields or persons at different hierarchy levels in a way they can
understand more easily. These conditions pose specific challenges for
requirements analysis, project management, training and support.

\subsection{RSE tasks and
responsibilities}\label{sec:tasks-and-responsibilities}

These skills, while already numerous are also generic on purpose. They
span a multidimensional space in which the day-to-day tasks and
responsibilities of an RSE can be found. We describe here some examples
of the competencies applied in combination to the set of current common
tasks and challenges for RSEs identified during the deRSE23 Paderborn
workshop.

The most obvious task of an RSE is to develop software that is used in
research. This broad topic requires all the \ac{SE} skills. Of course,
these are the competencies that are the most fluid since they have to
adapt to frequent technological advancements. Additionally, proper
\ac{SE} skills often require knowledge of \gls{TEAM}, and \gls{PM}.
Today, this means effective use of \acp{IDE}, \gls{static-analysis}
tools, \glspl{design-pattern} and documentation (for oneself and
others).

The RSE needs to be able to formulate and discuss structural and
behavioural aspects of software on a more general level than through the
code itself and often even before a first line of code is written. A set
of tools and diagrams for effective and standardised communication about
software on a meta level is provided by the \acp{UML}. As a modelling
tool, it is directly related to \gls{MOD}. Additionally, it can be
applied in various stages of the \gls{SWLC}, especially in the early
stages, as a first documentation of the planned modular structure to
facilitate \gls{DOCBB}, \gls{USERS}, \gls{PM} and \gls{TEAM}.

The RSE needs to be able to choose appropriate algorithms and techniques
(\gls{MOD} and \gls{NEW}). Apart from the technical feasibility, this
choice is also informed by the values outlined in \cref{sec:values}. For
example, the RSE needs to be able to estimate resource usage
(processing, memory and storage consumption,
e.g.~\autocite{Lannelongue2021}). Resource usage has not only a direct
financial price tag but also environmental costs via associated energy
consumption (see \cref{sec:environmental-sustainability}).

Software development also includes testing. This task is a manifestation
of the \ac{SE} competencies of \gls{DOCBB} and \gls{MOD} since a model
of the software is required in order to write good tests that facilitate
understanding and documentation. Today this encompasses the knowledge of
testing frameworks as well as \ac{CICD} practices. In addition to being
tested, software should also provide reproducible outputs. Projects like
ReproHack \autocite{ReproHack} can greatly help in fostering that
competency.

Apart from testing, there are many code analysis tools to monitor and
improve the quality of code. An RSE should be familiar with the tools
available for their specific environment and how to include some of them
into a \ac{CICD} pipeline. Typically, this includes linters and similar
static tools as well as dynamic tools like profilers and code coverage
analysis. The development of these tools is very dynamic and environment
specific. A good introduction can be found in~\autocite{swebok_2014} and
an online resource is~\autocite{vihps}. As these tools help with
behavioural and structural analysis and therefore modularisation these
tools enable \gls{MOD} as well as \gls{DOCBB}.

Part of the \ac{FAIR} principles is to make software findable and
reusable. The RSE needs to be able to decide when and why to keep a
repository private. This decision requires knowledge in \gls{RC},
\gls{USERS}, \gls{TEAM}, and sometimes \gls{SP}. Furthermore, knowledge
of the practices and contractual regulations of the RSE's institution is
also required.

The RSE also needs to understand metadata for research data and research
software. There are ongoing efforts on metadata for research software
such as CodeMeta \autocite{jones_codemeta_2017} and the \ac{NFDI}
working group \autocite{castro_research_sw_metadata_2023} on the
subject. These are complemented by the development of new tools and
methods for providing and working with software metadata, such as the
Citation File Format project~\autocite{druskat_cff_2021} and
HERMES~\autocite{druskat_hermes_2022}. Other efforts focus on Software
Management Plans
(e.g.,~\autocites{alves_elixir_2021}[~][]{martinez_ortiz_practical_2022})
which could be helpful for RSEs at early stages (i.e., with not much
experience of project management). They give quick hints on what to look
for regarding basic management for research software (including
information on, e.g., licenses, releases, publication, citation,
archiving) together with some ongoing work on corresponding
metadata~\autocite{giraldo_metadata_2023}. Metadata can also be used
actively during and within a research project, to inform the
decision-making processes~\autocite{Bird2016}.

Most RSEs will contribute to other projects, some of which will be
large. This is a topic that requires competency in \gls{SWREPOS},
\gls{SRU}, and \gls{SP} in order to understand the ramifications of
sharing, and in \gls{DOCBB}, since the contributed code has to be
understood by others. Interacting with project members depends on the
\gls{TEAM} skill. Today, this frequently involves the effective use of
collaborative platforms like \gls{GitHub}/\gls{GitLab}, honouring a
project's code of conduct, and some knowledge of popular open source
software licences, e.g.~the \ac{GPL}. The \gls{TEAM} skill will play a
major role when an RSE is introduced to an existing project. An existing
project will have grown some idiosyncratic habits and processes. Often
it will require all the skills and patience of an RSE to steer a project
towards best software engineering practices, while not having a
leadership position.

RSEs are embedded in communities. There are two different aspects to
finding these communities: First, we have the aspect of community
building for a research project. Since this deals with software that is
supposed to be used in research this requires knowledge of \gls{RC},
\gls{USERS}, and also \gls{NEW}, in order to effectively interact with
domain scientists. Today, an example is a presence on social media. The
other \gls{TEAM}-related aspect is the embedding of recently-trained
RSEs into the Research Software Engineering community, sharing the same
set of values and competencies. We envision newcomers to the RSE field
becoming part of a strong network of RSEs, tool-related communities, and
the classical domain communities, making them more effective at
supporting research. These networks are a lifelong manifestation where
RSEs work to provide an inclusive environment for their peers and
provide opportunities for lifelong learning. An ever-growing list of
national associations can be found at \autocite{RSECouncil}.

RSEs are also mentoring colleagues (see also
\cref{sec:mentoring-and-diversity}). This necessitates giving good
advice that fits to a project's stage in its life cycle, thereby
requiring knowledge of (\gls{SWLC}), and its context in its research
domain and thus (\gls{RC}).

Research software can often start out as a tool to answer a personal
research question, becoming more important when other researchers start
to rely on it. At the other end of the scale, research software can
sometimes underpin key processes that deal with critical questions such
as weather forecasting or medical diagnosis. A classification of
software is commonly used to formalise the process of giving good
advice~\autocite{Wang2012,Schlauch2018b} where research software can
move from one class to another during its life cycle.
\autocite{Schlauch2018b} classifies applications based on their scope
and criticality and provides \ac{SE} recommendations. The RSE needs to
be able to identify the application class they are dealing with and
apply the respective RSE practices.

Often RSEs, especially in RSE groups, will develop applications and
services with different variants for different research purposes and
groups. Additionally, many research groups develop their own codes for
specific research purposes, e.g.~simulation codes or specialised data
analysis pipelines. A lot of their development of new features is
project-based, often through PhD projects. Work can sometimes result in
code that diverges from the main project into a separate variant with
re-integration planned as a final step. To reduce the chance of variant
source code diverging significantly and producing a large integration
overhead, \gls{PM} skills and methods are needed. More specifically,
software product line management methods have been developed for this
exact problem and purpose.

\section{How much do different people need to
know?}\label{sec:how-much-to-know}

Now that we have the different competencies, we can explore various
dimensions of these competencies, depending on their circumstances. A
strong beneficiary of specialised RSEs can also be newly formed RSE
centres at research institutions.

\subsection{Career level}\label{career-level}

At different career levels, differing skills are required. To elaborate
on that, we have prepared the following tables with three levels of
experience in mind.

\begin{itemize}
\tightlist
\item
  Junior RSE: These are people who are in the earlier stages of their
  RSE career journey, but they should ideally have research experience
  of their own as well as the skills to contribute reliable and
  well-structured code to software projects.
\item
  Senior RSE: They have gained experience, both concerning their
  software skills as well as in their research collaborations in
  potentially many different fields. They can set the standards in a
  software project.
\item
  Principal RSE: Their actual job description varies a lot. These may be
  RSE team leaders based in a professional services type role, or they
  may be professors or research group leaders based in a more
  academic-focused role. They are often the people responsible for
  bringing in the funding that supports new and sustains existing
  projects. Generally speaking, they do not need to be actively involved
  in the day-to-day technical tasks, but they should be able to guide
  projects from both a technical and a research perspective while
  providing an inclusive working space.
\end{itemize}

\cref{tbl:comp-lvls-techn}, \cref{tbl:comp-lvls-res}, and
\cref{tbl:comp-lvls-comm} elaborate on the required facets of the
competencies in different roles. A story-like example of an individual
through the hierarchies can be found in \cref{subsec:examplecareer} .

\blandscape
\small
\renewcommand*{\arraystretch}{1.4}

\begin{longtable}[]{@{}
  >{\raggedright\arraybackslash}p{(\columnwidth - 6\tabcolsep) * \real{0.0909}}
  >{\raggedright\arraybackslash}p{(\columnwidth - 6\tabcolsep) * \real{0.3030}}
  >{\raggedright\arraybackslash}p{(\columnwidth - 6\tabcolsep) * \real{0.3030}}
  >{\raggedright\arraybackslash}p{(\columnwidth - 6\tabcolsep) * \real{0.3030}}@{}}
\caption{Levels of technical skills expected per RSE career stage.
\label{tbl:comp-lvls-techn}}\tabularnewline
\toprule\noalign{}
\begin{minipage}[b]{\linewidth}\raggedright
Competency
\end{minipage} & \begin{minipage}[b]{\linewidth}\raggedright
Junior RSE
\end{minipage} & \begin{minipage}[b]{\linewidth}\raggedright
Senior RSE
\end{minipage} & \begin{minipage}[b]{\linewidth}\raggedright
Principal RSE
\end{minipage} \\
\midrule\noalign{}
\endfirsthead
\toprule\noalign{}
\begin{minipage}[b]{\linewidth}\raggedright
Competency
\end{minipage} & \begin{minipage}[b]{\linewidth}\raggedright
Junior RSE
\end{minipage} & \begin{minipage}[b]{\linewidth}\raggedright
Senior RSE
\end{minipage} & \begin{minipage}[b]{\linewidth}\raggedright
Principal RSE
\end{minipage} \\
\midrule\noalign{}
\endhead
\bottomrule\noalign{}
\endlastfoot
\gls{SWLC} & Should be aware of the software life cycle. & Should know
where in the life cycle their project is and which decisions are likely
to lead to technical debt. & Should know how to manage and steer
development/project resources accordingly. Should also have an
understanding of the potential consequences of key project management
decisions. \\
\gls{DOCBB} & Should be able to write reusable building blocks. & Same
as junior, but the quality should set the standard for the project,
while following current best practices. & Should know the current best
practices and point their team members and collaborators to the right
resources. \\
\gls{DIST} & Should be able to use package distribution platforms. &
Same as junior, but should also be familiar with current best practices
for building and deploying packages. & Should ensure that their project
is available via an up-to-date and secure distribution platform. \\
\gls{SWREPOS} & Should seamlessly interact with the repository of their
project. & Should be well-versed in the intricacies and best practices
around working with a repository, and probably interact with
repositories of multiple projects. & Should promote the use of
repositories and be able to convey best practices of sharing and
reviewing to junior and senior RSEs. \\
\gls{MOD} & Should have a basic grasp of the part of the software they
are responsible for in order to use basic tools such as a debugger. &
Should understand the characteristics of large parts of the codebase
considering a variety of the metrics. & Should have a detailed
understanding of the software project as well as its aims and potential
for impact, in order to effectively steer it. \\
\end{longtable}

\newpage

\begin{longtable}[]{@{}
  >{\raggedright\arraybackslash}p{(\columnwidth - 6\tabcolsep) * \real{0.0909}}
  >{\raggedright\arraybackslash}p{(\columnwidth - 6\tabcolsep) * \real{0.3030}}
  >{\raggedright\arraybackslash}p{(\columnwidth - 6\tabcolsep) * \real{0.3030}}
  >{\raggedright\arraybackslash}p{(\columnwidth - 6\tabcolsep) * \real{0.3030}}@{}}
\caption{Levels of research skills expected per RSE career stage.
\label{tbl:comp-lvls-res}}\tabularnewline
\toprule\noalign{}
\begin{minipage}[b]{\linewidth}\raggedright
Competency
\end{minipage} & \begin{minipage}[b]{\linewidth}\raggedright
Junior RSE
\end{minipage} & \begin{minipage}[b]{\linewidth}\raggedright
Senior RSE
\end{minipage} & \begin{minipage}[b]{\linewidth}\raggedright
Principal RSE
\end{minipage} \\
\midrule\noalign{}
\endfirsthead
\toprule\noalign{}
\begin{minipage}[b]{\linewidth}\raggedright
Competency
\end{minipage} & \begin{minipage}[b]{\linewidth}\raggedright
Junior RSE
\end{minipage} & \begin{minipage}[b]{\linewidth}\raggedright
Senior RSE
\end{minipage} & \begin{minipage}[b]{\linewidth}\raggedright
Principal RSE
\end{minipage} \\
\midrule\noalign{}
\endhead
\bottomrule\noalign{}
\endlastfoot
\gls{NEW} & Should have some curiosity to fit into research teams. &
Same as junior, but they should proactively propose directions in
individual aspects of the project. & Should have research insights and a
broad view of the research field to steer the project. \\
\gls{RC} & Should be aware of the research life cycle. & Should know the
position of the project in the research life cycle. & Should know what
is necessary for the project to fit into its position in the research
life cycle. \\
\gls{SRU} & Should be aware of software reusability tools. & Should be
able to search with software reusability tools. & Should be able to
effectively search with \gls{SRU} tools and to evaluate and perform the
integration of a library into the project. \\
\gls{SP} & Should be aware of available opportunities to publish
software and understand the need to consider issues of intellectual
property. & Should be able to correctly publish software in simple cases
and to identify cases where professional legal advice is needed. & Same
as senior, plus the ability to take the future publication of software
into account when initiating and guiding larger software collaboration
projects. \\
\gls{DOMREP} & Should be able to interact with the domain repository. &
Same as junior RSE. & Same as junior, and should know about how it fits
into workflows surrounding these domain repositories. \\
\end{longtable}

\begin{longtable}[]{@{}
  >{\raggedright\arraybackslash}p{(\columnwidth - 6\tabcolsep) * \real{0.0909}}
  >{\raggedright\arraybackslash}p{(\columnwidth - 6\tabcolsep) * \real{0.3030}}
  >{\raggedright\arraybackslash}p{(\columnwidth - 6\tabcolsep) * \real{0.3030}}
  >{\raggedright\arraybackslash}p{(\columnwidth - 6\tabcolsep) * \real{0.3030}}@{}}
\caption{Levels of communication skills expected per RSE career stage.
\label{tbl:comp-lvls-comm}}\tabularnewline
\toprule\noalign{}
\begin{minipage}[b]{\linewidth}\raggedright
Competency
\end{minipage} & \begin{minipage}[b]{\linewidth}\raggedright
Junior RSE
\end{minipage} & \begin{minipage}[b]{\linewidth}\raggedright
Senior RSE
\end{minipage} & \begin{minipage}[b]{\linewidth}\raggedright
Principal RSE
\end{minipage} \\
\midrule\noalign{}
\endfirsthead
\toprule\noalign{}
\begin{minipage}[b]{\linewidth}\raggedright
Competency
\end{minipage} & \begin{minipage}[b]{\linewidth}\raggedright
Junior RSE
\end{minipage} & \begin{minipage}[b]{\linewidth}\raggedright
Senior RSE
\end{minipage} & \begin{minipage}[b]{\linewidth}\raggedright
Principal RSE
\end{minipage} \\
\midrule\noalign{}
\endhead
\bottomrule\noalign{}
\endlastfoot
\gls{TEAM} & Should be able to work in the team in order to effectively
fulfil the given tasks. Should be able to learn from code review. &
Should be able to break down tasks into more easily digestible sub-tasks
and review or guide work undertaken by less-experienced team members. &
Should be able to lead the team and set the respective direction. \\
\gls{TEACH} & Should be able to perform simple peer-to-peer on-boarding
tasks. & Should be able to explain logical components and the general
architecture to other RSEs. & Should be able to effectively communicate
about all high-level parts of the project. \\
\gls{PM} & Should be aware of the employed \gls{PM}~method. & Should be
able to use and adapt the employed \gls{PM}~method. & Should be able to
design and adapt the employed \gls{PM}~method. \\
\gls{USERS} & Should be able to communicate with both users and SEs on
the project, on topics of the research and SE. & Same as junior RSE, and
be able to interpret the feedback. & Same as senior, and should also be
able to effectively take feedback into account when steering the
project. \\
\end{longtable}

\elandscape

\subsection{Helpful RSE skills for researchers in an academic
career}\label{helpful-rse-skills-for-researchers-in-an-academic-career}

In the previous section, we looked at the competency levels needed for
RSE specialists. However, many of these competencies are important for
domain researchers in academia as well, who do not specialise in RSE but
nevertheless contribute to research software. Naturally, the `R'
competencies apply, and research in general is increasingly team based.
Additionally, many researchers in fields from classical examples like
numerical mathematics or theoretical physics to newer disciplines like
digital humanities will spend time in their research on writing and
developing software. Therefore, RSE focused training, e.g., in a
master's programme, is also beneficial for students in these fields
resulting in a broader audience. This also means that students as well
as researchers need to be given time to acquire those skills, e.g., to
be able to attend training in RSE-relevant topics as part of their
regular work or study.

This section outlines how the RSE competencies could be reflected at all
academic levels. Again, this relates to domain studies and non-RSE
positions in academia. It is important to note that this section does
not reflect the current state of academic training and research
institutions. Instead, it summarises the discussions with and between
workshop participants at different levels of academic progression on
what they would have liked to learn at an earlier stage or know before
starting their current position. While individuals already work at
implementing some of these changes and teaching these skills, it has not
yet reached a systemic level.

The text is organised along the academic progression path (bachelor's
degree, master's degree, PhD, Postdoc, \ac{PI}/Professor). Since each
level is based on the previous levels, we presume that the skills and
competencies at each level also encompass those of the previous levels.
Due to the broad need throughout academic specialisations, the described
levels serve as a baseline and certain fields will require higher
\ac{SE} skill levels as development is a large part of their actual
research.

\paragraph{Bachelor's level}\label{bachelors-level}

Students at the undergraduate level mostly consume science/knowledge.
During their studies, they should also learn about the existence of
digital tools and structures. Undergraduate students should be aware
that RSEs exist and that software has different quality aspects
(\gls{DOCBB}). They should be aware of domain specific tools
(\gls{DIST}, \gls{SRU}) and where to find them (\gls{SWREPOS},
\gls{DOMREP}). At this level, it may be sufficient to consider software
as black boxes (\gls{USERS}) although some training in data presentation
would be very helpful and a good way to find out about programming
(\gls{MOD}, \gls{NEW}). They should have a basic awareness of software
licences, such as legal pitfalls and implications for good scientific
practice (\gls{SP}). They will be taught about the research cycle
(\gls{RC}) and that researchers often work in groups (\gls{TEAM}).
During practicals, they will have an opportunity for peer learning
(\gls{TEACH}).

\paragraph{Master's level}\label{masters-level}

A student at a master's level can participate in science and should
therefore be able to use ``some'' digital structures. A master's student
needs to be aware of relevant tools and data sets for their domain,
where to find them and how to use them (\gls{DIST}, \gls{SWREPOS},
\gls{DOMREP}). They should be able to process and present their data
(\gls{MOD}). They need to understand how their research depends on
software (\gls{SWLC}). Working on their master's thesis allows them to
understand the research cycle (\gls{RC}), practice project management
(\gls{PM}) and collaborate with other members of their research group
(\gls{TEAM}).

\paragraph{PhD}\label{phd}

PhD students perform independent research under guidance. They need to
know relevant tools and structures. They should know where to find
information about tools and where to find help using them (\gls{DOCBB},
\gls{SWREPOS}). They should be able to use the tools (\gls{DIST}) and
identify and report bugs (\gls{MOD}). They need to be aware that the
user's perspective is different from the developer's perspective in
order to be able to write good bug reports (\gls{USERS}). They might
produce new software (\gls{MOD}, \gls{SRU}), in which case they need to
understand how to licence their code for publication (\gls{SP}). PhD
students need to be curious to be able to conduct their research. In
order to be able to explore new tools (\gls{NEW}) they must be able to
evaluate research software (\gls{SWLC}). They need to be able to
interact with services (\gls{RC}) and domain specific repositories
(\gls{DOMREP}). They should be able to supervise a student
(\gls{TEACH}).

\paragraph{Postdoc}\label{postdoc}

Postdocs are independent researchers. Their role is similar to that of a
PhD student, with a deepened focus on their research career. However,
they are proficient users of all relevant tools, which makes them active
contributors to their domain of research. They need to be aware of more
advanced topics regarding intellectual property rights, such as patents
(\gls{SP}).

\paragraph{PI/Professor}\label{piprofessor}

They are experts in their field and should be able to give proper
guidance to their students on which digital tools are currently
relevant. They should be aware of the skills of an RSE and when they
might need one in their group. They should encourage their students to
use relevant tools (\gls{DIST}). They need to be able to judge the
suitability of the software (\gls{SWLC}) and follow the interactions
between relevant projects (\gls{SWREPOS}). They should be able to advise
their students on the legal aspects of software production and
distribution (\gls{SP}). They should be able to contribute meaningfully
to the steering decisions of the software in their field (\gls{USERS}).
They are able to guide students and prepare and deliver a full lecture
course (\gls{TEACH}). They need to manage and lead their research group
(\gls{PM}, \gls{TEAM}).

\subsection{Project team structures}\label{project-team-structures}

In \cref{tbl:teams-se}, \cref{tbl:teams-res}, and \cref{tbl:teams-comm},
we look at individual or team competencies and approaches to them,
considering how these differ depending on whether an RSE is working
alone on a software project, or whether they are working as part of a
team of RSEs. We extend this to consider how things differ when an RSE
or a group of RSEs is based locally within a research team or
department, or when they are based in a dedicated, centralised RSE team.
We also look at organisational aspects in the context of each of the
considered competencies, since there are a variety of ways that
organisations can contribute to and support them, complementing those
proposed by~\autocite{Katerbow2018}. Some of them are brought to life in
the example career path of \cref{subsec:examplemaster}. We first
summarise the meaning of each of the columns in the tables:

\begin{itemize}
\tightlist
\item
  \textbf{Competency:} The code assigned to the competency being
  considered, as defined in \cref{sec:required-generic-skills},
  e.g.~\gls{TEAM}.
\item
  \textbf{Individual RSE (Locally-based):} A single person working on
  software within a research project - for example a domain RSE with
  focus on their own specific research. Often time-constrained, may be
  self-taught.
\item
  \textbf{Individual RSE (RSE team-based):} A single person working on
  research software - generally a professional RSE assigned to support
  another team's software on their own, who however is connected to an
  RSE team.
\item
  \textbf{Group of RSEs (Locally-based):} A group within a research
  group or team, working together on software to support or undertake a
  single research goal/project. Similarly to the individual RSE, they
  are often research-focused with RSE skills, often self-taught.
\item
  \textbf{Group of RSEs (RSE team-based):} An RSE team working together
  on research software projects for a research group.
\item
  \textbf{Organisation-level RSE support:} Describes how the defined
  competencies are recognised and represented at an organisational level
  and what the organisation can do to support the RSEs in the context of
  the different team structures. These can be read as policy/action
  recommendations.
\end{itemize}

These tables take the perspective of the expected skill set of each RSE
or team of RSEs, similarly to personas in a user experience analysis.
The current situation may differ.

\blandscape
\small
\renewcommand*{\arraystretch}{1.4}
\begin{longtable}{|p{1.8cm}|p{3.5cm}|p{3.5cm}|p{3.5cm}|p{3.5cm}|p{4.5cm}|}
    \caption{Levels of software eng. skills expected per team structure.}\label{tbl:teams-se}\\
    \hline
    \multirow{2}{*}{Competency} & \multicolumn{2}{c|}{Working as an individual RSE}
    & \multicolumn{2}{c|}{Working with a group of RSEs} & \multirow{2}{*}{Organisation-level support} \\
    \cline{2-5}
              & Locally-based & RSE-Team based & Locally-based & RSE-Team based &\\\hline
    %%%%%%%%%%%%%%%%%%%%%%%%%%%%%%%%%%%%%%%%%%%%%%%%%%%%%%%%%%%%%%%%%%%%%%%%%%%%
    \gls{DOCBB}&
    Focuses on supporting research. May not be very familiar with code quality and
      structure. Follows basic best practice guides.&
    Puts greater focus on reusability, documentation, and knowledge of best practices,
      but potentially lacks domain knowledge.&
    Has more opportunities to discuss and share ideas, but team members may be
      less aware of key practices.&
    Has stronger ingrained focus on team-based \gls{PM} and development
      methodologies, resulting in higher quality, more reusable code.&
    Should offer training and other resources in core topics to support individual RSEs.
      Should have research software guidance/policies that provide advice.\\
    \hline
    %%%%%%%%%%%%%%%%%%%%%%%%%%%%%%%%%%%%%%%%%%%%%%%%%%%%%%%%%%%%%%%%%%%%%%%%%%%%
    \gls{DIST}&
    Does not emphasise code reusability and sharing/distribution.&
    Puts greater focus on reusability/sharing, but likely not as part of the project aims.&
    May want to develop reusable, shareable outputs for a specific case. Needs clear guidelines.&
    Focuses on quality and best practices. Reusability/packaging driven by project needs and spec.&
    Should provide policies on reusability/sharing. May be driven by requirements/policies, e.g., of institution or funding agency.
    \\\hline
    %%%%%%%%%%%%%%%%%%%%%%%%%%%%%%%%%%%%%%%%%%%%%%%%%%%%%%%%%%%%%%%%%%%%%%%%%%%%
    \gls{SWLC}&
    Manages the complete life cycle, \glspl{bus-factor} equal to 1.&
    The team supports parts of the software life cycle, but with low bus factor.&
    The team infrastructure and tooling supports the life cycle and sustainability.&
    The bus factor may still be low in parts of the code.
      Need to think about coherent life cycle management across the team - generally
      a key area of expertise for an RSE team.&
    Should support with training. Organisation may also provide site
      licences for, e.g., management tools.
    \\\hline
    %%%%%%%%%%%%%%%%%%%%%%%%%%%%%%%%%%%%%%%%%%%%%%%%%%%%%%%%%%%%%%%%%%%%%%%%%%%%
    \gls{SWREPOS}&
    Uses repositories for code management and demonstrating outputs,
    e.g., for supporting academic credit, but may be missing skills.&
    As locally-based, but professional RSEs are generally very experienced
      with use of repositories and their many features.&
    Uses repositories to collaborate inside the team.
      Can benefit from short courses on effective use.&
    Uses repositories extensively for project
      management, issue tracking, etc. in addition to code itself.
      May train others.&
    Should offer enterprise repository set ups,
      site licences etc. Also provide training
      for this vital research software development tooling.
    \\\hline
    %%%%%%%%%%%%%%%%%%%%%%%%%%%%%%%%%%%%%%%%%%%%%%%%%%%%%%%%%%%%%%%%%%%%%%%%%%%%
    \gls{MOD}&
    Needs full awareness of entire codebase to extend/maintain.
      If project taken on from another developer,
      there may be challenges in transferring the mental model.&
    As local, but more aware of need for future transition to other
      RSE(s), likely provides docs, issues, and other
      support from central services to support this. May only need to know
      parts of the code.&
    Internal team training ensures ability to build necessary
      mental model of codebase and to document it via text or tools for
      sustainability.&
    As local team, but likely more aware of tooling and practices
      in place within RSE team. Distributing work makes it only necessary
      for each developer to understand code related to their assigned tasks.&
    Should provide training and retain experience via
      coordinating and provide support for mentoring/community activities.
      Establishing RSE departments with specialists for certain aspects of software
      will improve overall turnaround times.
    \\\hline
    %%%%%%%%%%%%%%%%%%%%%%%%%%%%%%%%%%%%%%%%%%%%%%%%%%%%%%%%%%%%%%%%%%%%%%%%%%%%
\end{longtable}

\begin{longtable}{|p{1.8cm}|p{3.5cm}|p{3.5cm}|p{3.5cm}|p{3.5cm}|p{4.5cm}|}
    \caption{Levels of research skills expected per team structure.}\label{tbl:teams-res}\\
    \hline
    \multirow{2}{*}{Competency} & \multicolumn{2}{c|}{Working as an individual RSE}
    & \multicolumn{2}{c|}{Working with a group of RSEs} & \multirow{2}{*}{Organisation-level support} \\
    \cline{2-5}
              & Locally-based & RSE-Team based & Locally-based & RSE-Team based &\\\hline
    %%%%%%%%%%%%%%%%%%%%%%%%%%%%%%%%%%%%%%%%%%%%%%%%%%%%%%%%%%%%%%%%%%%%%%%%%%%%
    \gls{NEW}&
    May struggle to learn new methods and skills due to split research focus
      between research goal and software project.&
    Gets support from the RSE team to explore new methods and skills,
      make relevant contacts and learn more about the domain.&
    Has increased interest in learning new methods and skills,
      but still prioritises domain research.&
    As team-based individual&
    Should reach out to relevant local groups to facilitate training
      and sharing of know-how on new technical processes and tooling.
    \\\hline
    %%%%%%%%%%%%%%%%%%%%%%%%%%%%%%%%%%%%%%%%%%%%%%%%%%%%%%%%%%%%%%%%%%%%%%%%%%%%
    \gls{RC}&
    Is familiar with the \gls{RC} in their domain,
      especially when embedded in a research team.&
    Is familiar with the \gls{RC},
      although they may not have domain knowledge, which a group can provide.&
    Is familiar with the \gls{RC} and can share knowledge within the team.&
    One or more members of the team are strongly aware of the
      \gls{RC}.&
    Should provide extensive infrastructure to manage
      the \gls{RC}, supporting researchers/RSEs.
    \\\hline
    %%%%%%%%%%%%%%%%%%%%%%%%%%%%%%%%%%%%%%%%%%%%%%%%%%%%%%%%%%%%%%%%%%%%%%%%%%%%
    \gls{SRU}&
    Has limited awareness of existing solutions and limited support regarding \gls{SRU}.&
    Is familiar with software sharing and can discover tools and platforms.&
    As locally-based individual, but being part of a team can help to address this.&
    As team-based individual&
    Should run local environments to host software, catalogue software,
      and/or provide institution-level access to platforms that support this.
    \\\hline
    %%%%%%%%%%%%%%%%%%%%%%%%%%%%%%%%%%%%%%%%%%%%%%%%%%%%%%%%%%%%%%%%%%%%%%%%%%%%
    \gls{SP}&
    Has limited knowledge and motivation regarding \gls{SP}.&
    Applies practices, workflows, and policies established in the RSE team.&
    As locally-based RSE&
    As team-based RSE&
    Should raise awareness about software as a publishable scientific output, provide recommendations and checklists to support software publications, and have legal experts in place to offer advice on complex cases.
    \\\hline
    %%%%%%%%%%%%%%%%%%%%%%%%%%%%%%%%%%%%%%%%%%%%%%%%%%%%%%%%%%%%%%%%%%%%%%%%%%%%
    \gls{DOMREP}&
    Domain researchers working on software are likely to be more familiar
      with the domain-specific solutions.&
    RSEs may need guidance from domain researchers around domain-specific
      repositories if they have a background in a different domain.&
    As locally-based individual&
    As team-based individual&
    Should host domain-specific repositories for areas that the organisation works
      extensively in, but this is likely to be handled at a research group level.
    \\\hline
  \end{longtable}

\newpage

\begin{longtable}{|p{1.8cm}|p{3.5cm}|p{3.5cm}|p{3.5cm}|p{3.5cm}|p{4.5cm}|}
    \caption{Levels of communication skills expected per team structure.}\label{tbl:teams-comm}\\
    \hline
    \multirow{2}{*}{Competency} & \multicolumn{2}{c|}{Working as an individual RSE}
    & \multicolumn{2}{c|}{Working with a group of RSEs} & \multirow{2}{*}{Organisation-level support} \\
    \cline{2-5}
              & Locally-based & RSE-Team based & Locally-based & RSE-Team based &\\\hline
    %%%%%%%%%%%%%%%%%%%%%%%%%%%%%%%%%%%%%%%%%%%%%%%%%%%%%%%%%%%%%%%%%%%%%%%%%%%%
    \gls{USERS}&
    May have additional skills to safeguard potential future development
      and maintenance of the software for external users. Resourcing for future maintenance may be a challenge.&
    Has additional skills or can access support to safeguard potential
      future development and maintenance of the software for external users.&
    Needs to safeguard future development and maintenance of the software for
      external users, but may not have the skills or resources to support this.&
    Applies best practices to prepare the code for external users,
      while the team provides infrastructure and/or specialised RSEs for user support.&
    Should have institutions that are able to offer support with outreach and publicising outputs.
    \\\hline
    %%%%%%%%%%%%%%%%%%%%%%%%%%%%%%%%%%%%%%%%%%%%%%%%%%%%%%%%%%%%%%%%%%%%%%%%%%%%
    \gls{TEACH}&
    May be independently involved in training activities.&
    May be able to support researchers with core technical skills.&
    Shares knowledge and skills within the group (peer support).&
    Supports teaching more widely, either through organised courses
      or ad hoc activities such as ``code clinics''.&
    Should have programs for a diverse range of teaching/training activities,
      such as an RSE curriculum, as described in~\autoref{subsec:examplemaster}.
    \\\hline
    %%%%%%%%%%%%%%%%%%%%%%%%%%%%%%%%%%%%%%%%%%%%%%%%%%%%%%%%%%%%%%%%%%%%%%%%%%%%
    \gls{PM}&
    Is organised enough to be able to transfer the codebase to future RSEs.&
    Follows the project management approach set by the team, or can suggest such \gls{PM} approaches.&
    Has additional \gls{PM} challenges, but may not have awareness of or experience with key \gls{PM} skills,
      which can be acquired with low-key courses.&
    Team provides well-structured approaches and tooling to support management of projects.&
    Should offer training to support management of projects.
      May offer organisation-level tooling.
    \\\hline
    %%%%%%%%%%%%%%%%%%%%%%%%%%%%%%%%%%%%%%%%%%%%%%%%%%%%%%%%%%%%%%%%%%%%%%%%%%%%
    \gls{TEAM}&
    When not developing code for themselves, they must be able to work effectively with researchers they are potentially developing code for.&
    Must be able to work effectively with their home RSE team, as well as
      with researchers they are potentially developing code for.&
    Must have strong team skills and knowledge
      to support team-based software development.&
    Must be able to work and collaborate effectively in an interdisciplinary team,
      use required tools and processes, infrastructure, etc.&
    Should offer support with team work and promote interdisciplinary interaction. Should facilitate team-building initiatives,
    also on a social level.
    \\\hline
\end{longtable}
\elandscape

In the tables above, we have looked at how different competencies can be
related to and handled by researchers and RSEs working in different
environments within an organisation and how the organisations themselves
can contribute. We recognise that this is a challenging area to gain a
detailed view of and that this is still a significant generalisation. We
talk about the ``Research Software Engineer'' as a single entity but as
the field expands, we expect to see more roles and job titles emerging
around the RSE concept, many of which fit under the wider umbrella of
research technology professionals (RTPs)~\autocite{ukri_rtp}
\autocite{techniciancommitment}. Examples are different RSE-like
computational roles of the \ac{EMBL-EBI} BioExcel competency
framework~\autocite{BIOEXCEL} (also \cref{subsec:emblbio}), as is a
range of different roles from King's Digital Lab at King's College
London~\autocite{KDL}.

\section{RSE specialisations}\label{sec:rse-specialisations}

What we have defined above is intended to be a set of base skills that
an RSE irrespective of domain, position, and experience should know
about. There is a large variety of RSEs. They specialise in different
areas, some of which we want to present below. Many of the
specialisations may overlap, so the same RSE might for example work on
data management and open science. We categorise them into those that can
be viewed as a specialisation within RSE-specific topics, while other
RSEs might expand their skill set and profession to areas that are not
typical for an RSE.

\subsection{Specialisations within the core RSE
competencies}\label{specialisations-within-the-core-rse-competencies}

\paragraph{Open science RSE}\label{open-science-rse}

Open science and FAIRness of data and software are increasingly
important topics in research, as exemplified by the demand of an
increasing amount of research funding agencies requiring openness.
Hence, an open science RSE is required to have a deeper knowledge of
(\gls{RC}) and how to distribute software publicly (\gls{SRU},
\gls{SP}). Open Science RSEs can help researchers navigate the technical
questions that come up when practising Open Science, such as ``How do I
make my code presentable?'', ``How do I make my code citable?'', ``What
do I need to do to make my software \ac{FAIR}?'', or ``How do I
sustainably work with an (international) team on a large code base?''.
Like the Data-focused RSE, they have a deep understanding of \ac{RDM}
topics.

\paragraph{Project/community manager
RSEs}\label{projectcommunity-manager-rses}

When research software projects become larger, they need someone who
manages processes and people. In practice, this concerns change
management for code and documentation and community work to safeguard
usability and adaptability, but also handling project governance and
scalable decision-making processes. This gap can be filled by people who
invest in the (\gls{PM}), (\gls{USERS}), and (\gls{TEAM}) skills, as
exemplified in \cref{subsec:examplecareer}. Building a community around
a research project is an important building block for sustainable
software~\autocite{Segal2009}, so these RSEs play an important role,
even if they do not necessarily touch much of the code themselves.

\paragraph{Teaching RSEs}\label{teaching-rses}

RSEs interested in developing their (\gls{TEACH}) skill can focus on
teaching the next generation of researchers and/or RSEs and will play a
vital role in improving the quality of research software. They need to
have a good understanding of all RSE competencies relevant to their
domain and additionally should have teaching experience and training in
didactics and pedagogy.

\paragraph{User interface/user experience designers for research
software}\label{user-interfaceuser-experience-designers-for-research-software}

Scientific software is a complex product that often needs to be refined
in order to be usable even by other scientists. To facilitate this,
there are people required that specialise in the (\gls{DOCBB}) and
probably the (\gls{DIST}) competency with a focus on making end-user
facing software really reusable and hence \ac{FAIR}. This task is
supported by strong (\gls{MOD}) skills to reason about the behaviour of
potential users of the software.

\subsection{Specialisations outside the core RSE
competencies}\label{specialisations-outside-the-core-rse-competencies}

\paragraph{\$\{DOMAIN\}-RSE}\label{domain-rse}

While software is the common focus of all RSEs, there will be RSEs that
have additionally specialised in the intricacies of one particular
research domain, such as medical RSEs, digital humanities RSEs, or
physics RSEs. This can often serve as a base domain for RSE
specialisation as in \cref{subsec:examplemaster}.

\paragraph{Data-focused RSE}\label{data-focused-rse}

Data-focused RSEs work at the flourishing intersection between data
science and RSE. They are additionally skilled in cleaning data and/or
running data analyses and can help researchers in setting up their
analysis pipeline and/or \ac{RDM} solutions. When the field requires
research on sensitive data or information, e.g., patient information in
medicine, this RSE should have knowledge about secure transfer methods
and/or ways to anonymise the data. As part of \ac{RDM}, this RSE profile
is able to support all stages of the research data life
cycle~\autocite{Nieva2020}, with synchronous data management processes.
Those processes implement established best practices for planning and
documenting of data acquisition in a \ac{DMP}, as well as for
management, storage, and preservation of data, and publication and
sharing of data in repositories according to the \ac{FAIR}
principles~\autocite{FAIR}.

\paragraph{Research infrastructure
RSE}\label{research-infrastructure-rse}

This RSE has a special interest in \glspl{SysOp} and system
administration and sets up \ac{IT} infrastructures for and with
researchers. Therefore, this specialisation on the one hand requires a
deep knowledge of physical computer and network hardware and on the
other hand knowledge about setup and configuration of particular server
software, e.g., setup of virtual machines on hypervisors or the planning
and setup of compute server clusters for special purposes, e.g.,
\ac{ML}. As an interface between the researchers and the infrastructure,
they take care of user management, access permissions, and configuration
of required services.

\paragraph{HPC-RSE}\label{hpc-rse}

RSEs with a focus on \ac{HPC} have specialist knowledge about
programming models that can be used to efficiently undertake large-scale
computations on parallel computing clusters. They may have knowledge of
(automatic) code optimisation tools and methods and will understand how
to write code that is optimised for different types of computing
platforms, leveraging various efficiency related features of the target
hardware. They are familiar with \acrshort{HPC}-specific package
managers and can build dependencies from sources. They also understand
the process of interacting with job scheduling systems that are often
used on \ac{HPC} clusters to manage the queuing and running of
computational tasks. \acrshort{HPC}-focused RSEs may be involved with
managing \ac{HPC} infrastructure at the hardware or software level (or
both) and understand how to calculate the environmental impact of
large-scale computations. Their knowledge of how to run \ac{HPC} jobs
and write successful \ac{HPC} access proposals can be vitally important
to researchers wanting to make use of \ac{HPC} infrastructure.

\paragraph{ML-RSE}\label{ml-rse}

The development of research software based on \ac{ML} requires
additional specialised theoretical background and experienced handling
of appropriate software in order to produce meaningful results. This
involves knowledge about data analysis and feature engineering, metrics
that are involved in \ac{ML}, \ac{ML} algorithm selection and cross
validation, and knowledge in mathematical optimisation methods and
statistics. Here, we use \ac{ML} in a broad sense of machine-based
learning including deep learning, reinforcement learning, neuro-symbolic
learning and similar.

ML-RSEs analyse and check the suitability of an algorithm. They check if
it fulfils the needs of a certain task and they play a central role in
deciding on and selecting \ac{ML} libraries for a given task. The
increasing usage of \ac{ML} in numerous scientific areas with social
impact involves an emphasised awareness and consideration of possible
influences and biases. At the intersection of data
science~\autocite{SSIDataScience} and data-focused RSEs, the complex way
of solving problems utilising \ac{ML} calls for this separate
specialisation.

\paragraph{Legacy RSEs}\label{legacy-rses}

Research software may have evolved over generations of researchers
without change management or governance processes, while software
``ecosystems'' (e.g., programming languages, frameworks, operating
systems) constantly evolve. This may lead to the emergence of legacy
code that is still actively used. To safeguard continued usability and
adoption, these RSEs have experience in working with code written in
language standards and on software stacks considered deprecated by their
communities. Adaption of existing, large-scale codebases to evolving
dependencies (\gls{DIST}) or changing hardware (\ac{HPC}; see the
HPC-RSE specialisation) may require mastery in refactoring techniques
and in the usage of specialised code transformation tools.

\paragraph{Web-development RSE}\label{web-development-rse}

This RSE is skilled in the development of web applications and/or mobile
apps. They have expertise in one or more of frontend development,
backend development and the design or implementation of APIs, for
example to support research data portals or big research projects. Since
a lot of web services for research may be accessible to a large audience
or even to the public, this RSE is also familiar with aspects relating
to cybersecurity, usability and accessibility. Not only do they need to
balance these concerns while adhering to their values from
\cref{sec:values}, but they also need to efficiently communicate the
decisions made to stakeholders.

\paragraph{Legal-RSE}\label{legal-rse}

RSEs are often the go-to person for questions about software licensing,
in particular when mixing software components that use different
licences. But with the rising requirements from legislation, we foresee
the need for RSEs that still have a background in RSE but extend it with
a knowledge of legal processes that cover corner cases and go beyond
applying Best Practice guides. These requirements may arise in the area
of publication of research software, as this also requires knowledge
about particular laws or regulatory frameworks concerning data
protection, like the \ac{GDPR} within the \ac{EU}~\autocite{GDPR}.
Another area are legal aspects of cybersecurity and export control in
science and research (see~\autocite{ExportControl} for Germany).
Legal-RSEs focus on facilitating the achievement of technically feasible
solutions, while adhering to regulatory mandates. They are able to
communicate and collaborate effectively with lawyers.

\section{Future work}\label{sec:future-work}

This list and description of competencies is a first step to finding
common ground around which to structure curricula, institutions, and
teachers in this framework. Applications of these competencies in an
individual's career can be found in~\cref{subsec:examplemaster}.
Opportunities for sustainable funding is a concern that is often
\autocite{Goble2014,RSESofN2017,Carver2021,Mundt2022} raised by RSEs. An
omission that we found and that we would like to highlight in order to
spark a community discussion is that RSEs that choose explicitly a
science-supporting role outside of research will not be eligible for
funding under the statutes of many funding organisations that require at
least a PhD.

To alleviate this and to give RSEs in leadership positions a means to
become eligible for funding themselves, since completion of scientific
training is often a requirement~\autocite{DFG_50_01}, we see two
possible parts of a solution. One is to allow for doctorates primarily
based on software contributions to the scientific community. Secondly,
we propose the introduction of new, standardised certificates like those
in~\cref{subsec:examplecareer}, and to officially accept them as
PhD-equivalent concerning eligibility to be a \ac{PI}. Beyond this
discussion, a diverse set of publications on the topic RSE teaching is
already in the making.

Within this set, we will work next on how to institutionalise education.
In that publication, we will detail how we organise our institutions and
what qualifications our teachers need to have in order to effectively
communicate our values. We will put forward ideas on how to build up
bachelor's and master's programmes, of which a glimpse can already be
found in \cref{subsec:examplemaster}. We will show how we intend to
provide the necessary continuous education for RSEs after graduation,
and we will connect that with the integration of RSEs into a mesh of
community networks aimed at supporting research, while providing them
with an inclusive social network that further facilitates lifelong
learning. That publication will again intentionally be free of regional
specifics, to also serve as a blueprint that other national RSE
societies can build upon.

Online resources for courses are another important building block. This
is the general intention of the learn-and-teach
project~\autocite{learnandteach}. Surveying and curating of existing
resources is not carried out as a traditional publication, but it is
made available as a continuously-evolving online resource
at~\autocite{learnandteach}.

And finally, we plan to formulate a call to action, building on the
previously mentioned publication on the necessary institutions, that
spells out everything that is required to best support the continuous
need for young RSEs to support digital science specifically in Germany.

\section{Conclusion}\label{sec:conclusion}

This paper started from a community workshop at deRSE23 in Paderborn
where people working in RSE related fields got together to figure out
structures and ideas for educating newcomers to this field. One outcome
of this diverse gathering is that RSEs from differing fields gather
around similar core concepts, At the same time they share a vision of
how to renew scientific research practice making extensive use of
digital tools. In this publication, we have tried to formalise these
concepts. We have formulated a set of values that guide our actions in
society, manifestly making RSEs part of the scientific community that
shares the ideals of good scientific practice. At the same time, being
close to software engineers, we cherish that we have to take
responsibility for our tools. We listed core competencies that have been
intentionally formulated abstractly without referencing any particular
information-processing device. As expected, we have drawn equally upon
notions from \ac{SE} and other research fields, but found that we
likewise require teamwork capabilities. We detailed these competencies
in various dimensions and found that a different amount is required in
different positions and scientific domains. Using this, we proposed
recommendations for organisations to foster the development of these
competencies.

The gathered values and competencies form a common denominator that
unifies RSEs and enables them to identify with this domain, in the
knowledge that it is already or will soon become critically important
for many areas of science. These competencies at the intersection of
research and SE, coupled with a firm belief in team processes, make RSEs
sought after on the job market and their values make them responsible
members of a digital society. The result is a qualification profile
which is highly attractive for young people.

At an institutional level, research performing organisations have a
growing interest in fostering RSE training to support the use of
\ac{FAIR} data and \ac{FAIR} software in the academic world, a direction
determined by new incentives created by scientific journals and
librarians. How we update existing institutions and set up new ones that
provide this education will be the topic of a follow-up paper.

\section*{Contribution details}\label{contribution-details}
\addcontentsline{toc}{section}{Contribution details}

Heidi Seibold came up with the original idea for the deRSE23 workshop in
Paderborn. Heidi Seibold, Jeremy Cohen, Florian Goth, Renato Alves, Jan
Philipp Thiele, and Samantha Wittke organised the deRSE23 workshop. We
thank all the participants of this community workshop! Toby Hodges
conceptualised and organised the un-deRSE23 workshop together with Jan
Philipp Thiele and Florian Goth. We also thank all the participants of
this follow-up community workshop! Jeremy Cohen, Gerasimos Chourdakis,
Magnus Hagdorn, Jean-Noël Grad, Jan Philipp Thiele, and Matthias Braun
organised the deRSE24 workshop in Würzburg. We are also grateful to the
participants of this third community workshop! Heidi Seibold, Jeremy
Cohen, Florian Goth, Renato Alves, Jan Philipp Thiele, Jan Linxweiler,
Jean-Noël Grad, and Samantha Wittke contributed the initial draft.
Florian Goth supervised the project and did the project administration.
Jean-Noël Grad designed and implemented the software tooling for the
collaborative writing of this manuscript on GitHub. Everybody
contributed to the final review and editing.

The CRediT system \autocite{Brand2015} is far too generic to adequately
describe the contributions of everybody in various workshops, spread
over a two year period. While everybody contributed to the discussion
formulating and refining the ideas, and to collaboratively writing
and/or reviewing and editing the entirety of the script, some parts
merit special mention. Renato Alves quickly jumped in to host the first
deRSE23 workshop to take over from a sick organiser. Matthias Braun
contributed early versions of the specialisations and also contributed
to the survey. Leyla Jael Castro contributed to the initial draft of the
example career path, and provided helpful insights in discussions on
metadata. Gerasimos Chourdakis' contributions to the paper are numerous
(extensively editing large parts of the initial draft), but he
especially wrote first drafts for clarifying the relationship of the RSE
competencies to the SE competencies. He also designed the ``Learning and
teaching RSE'' website \autocite{learnandteach}. Simon Christ helped
with typesetting and contributed the competencies' symbols and the
spell-checker script. Jeremy Cohen drafted the initial introduction and
contributed the tables for RSEs in centralised RSE departments. Stephan
Druskat contributed parts on proper software citation and publication
and sharpened various RSE specialisations. Fredo Erxleben contributed to
early discussions of the paper and added the contributions by the
Helmholtz Association. Jean-Noël Grad contributed initial drafts for the
ELIXIR framework and the section on the work of the HPC certification
forum as well as numerous other contributions to the BibTeX
infrastructure and the GitHub actions. Magnus Hagdorn drafted and
supervised the ethics and values section for an RSE and made sure that
these values are reflected in the competencies of RSEs. Toby Hodges
contributed parts on the Carpentries, and helped steer the curriculum
discussion. Guido Juckeland contributed experiences from his first RSE
course for students. Dominic Kempf drafted the first version of the
example curriculum. Anna-Lena Lamprecht helped with proper wording,
especially with awareness about established SE terminology, that was
misused earlier. Jan Linxweiler drafted various RSE specialisations and
made sure that clean coding techniques got their due recognition. Frank
Löffler rewrote numerous parts to be actually legible, and helped with
preparation for the final steps of a de-RSE position paper. Michele
Martone wrote the first draft of the environmental sustainability
section. Moritz Schwarzmeier drafted the categorisation of the
specialisations. Heidi Seibold contributed the idea and started
everything. Jan Philipp Thiele drafted initial parts of the technical
pillar of the RSE competencies, and represented the project on numerous
discussions. Harald von Waldow contributed to initial drafts of the
Masters program and contributed his knowledge to the explainability of
computer simulations. Samantha Wittke contributed the parts on
CodeRefinery and how to reach out to new RSEs. Florian Goth has the
pleasure of being grateful to all collaborators in this project for
contributing their time and knowledge into this project!

\section*{Acknowledgements}\label{sec:acknowledgements}
\addcontentsline{toc}{section}{Acknowledgements}

FG acknowledges funding from the Deutsche Forschungsgemeinschaft (DFG,
German Research Foundation) through the SFB 1170 ``Tocotronics'',
project Z03 - project number \geprislink{258499086} as well as financial
support by the Deutsche Forschungsgemeinschaft (DFG, German Research
Foundation) under Germany's Excellence Strategy through the
Würzburg-Dresden Cluster of Excellence on Complexity and Topology in
Quantum Matter -- ct.qmat (EXC 2147, project-id \geprislink{390858490}).

MB acknowledges support by the Deutsche Forschungsgemeinschaft (DFG,
German Research Foundation) under Germany's Excellence Strategy -- EXC
2120/1 -- \geprislink{390831618}.

LJC acknowledges support from the NFDI4DS consortium funded by the
German Research Foundation (DFG) - project number
\geprislink{460234259}.

JC acknowledges support from the UK Engineering and Physical Sciences
Research Council (UKRI-EPSRC) under grants EP/R025460/1, EP/W035731/1
and EP/Y530608/1.

JNG acknowledges funding from the Deutsche Forschungsgemeinschaft (DFG,
German Research Foundation) - project number \geprislink{391126171} (PI:
Holm), from the German Federal Ministry of Education and Research
(Bundesministeriums für Bildung und Forschung, BMBF) under the funding
code
\href{https://www.kooperation-international.de/foerderung/projekte/detail/info/verbundprojekt-multixscale-hpc-exzellenzzentrum-fuer-multi-skalen-simulationen-auf-hoechstleitungsrechnern-1}{16HPC095},
and from the European Union -- this work has received funding from the
European High Performance Computing Joint Undertaking (JU) and countries
participating in the project under grant agreement No
\href{https://doi.org/10.3030/101093169}{101093169}.

DK acknowledges support from the Scientific Software Center which is
funded as part of the Excellence Strategy of the German Federal and
State Governments.

MM acknowledges funding from the SiVeGCS Project.

MS would like to thank Hessian Ministry of Higher Education, Research,
Science and the Arts and the Federal Government and the Heads of
Government of the Länder, as well as the Joint Science Conference (GWK),
for their funding and support within the framework of the NFDI4Ing
consortium. Part of this work was funded by the German Research
Foundation (DFG) - project number \geprislink{442146713}. Part of this
work was funded by the Hessian Ministry of Higher Education, Research,
Science and the Arts - cluster project Clean Circles.

We appreciate the comments and suggestions from Yves Vincent Grossmann,
Wilhelm Hasselbring, and Bernhard Rumpe.

\appendix

\section{Appendix}\label{sec:appendix}

\subsection{An example master's programme for research software
engineering}\label{subsec:examplemaster}

The target audience for such a master's programme are students holding a
bachelor's degree from a domain science, which we will call ``home
domain'' in the following. There is explicitly no restriction on the
candidates' home domain: it may be from the \ac{STEM} disciplines, life
sciences, humanities or social sciences, and it can also change later in
their career. Candidates with a bachelor's degree in computer science
are also explicitly included, although we acknowledge that their
master's programme should include adaptations to make their interaction
effective with other domain scientists. In order to give the future RSE
the necessary breadth, we expect this to be a four-semester curriculum.

The curriculum is formed from a combination of modules, some of which
are core modules teaching essential skills that must be completed by all
students. Other modules introduce more specialised concepts and skills.
During the master's programme, students should pick an RSE
specialisation from the list in this paper and attend these additional
modules to deepen their knowledge in that field.

Core modules are of course drawn from the three pillars of the RSE and
can be categorised accordingly.

\begin{itemize}
\tightlist
\item
  Software/Technical skills:

  \begin{itemize}
  \tightlist
  \item
    Foundational module: Here we have an introduction to programming:
    Emphasising use cases over programming paradigms, students learn at
    least two languages: a language that facilitates prototyping and
    data processing (e.g., \gls{Python} or \gls{R}) and a language for
    designing complex, performance-critical systems (e.g.,
    \gls{C}/\gls{Cpp}). This exposes them to computers in a hands-on
    fashion and is the foundation for (\gls{DOCBB}, \gls{DIST}).
  \item
    Computing environment module: Programming languages are not enough
    to work in a landscape of many interconnected software components;
    hence we require something like software craftsmanship, where tools
    such as the Unix shell, version control systems, build systems,
    documentation generators, package distribution platforms, and
    software discovery systems are taught to strengthen skills in
    (\gls{DIST}, \gls{DOCBB}, \gls{SWREPOS}, \gls{SRU}).
  \item
    Software engineering module: Here we develop foundational software
    engineering competencies (basic knowledge and skill regarding
    requirements engineering, software architecture and design,
    implementation, quality assurance, software evolution), again
    emphasising and strengthening (\gls{DOCBB}, \gls{DIST}) on a more
    abstract level.
  \end{itemize}
\item
  Research skills:

  \begin{itemize}
  \tightlist
  \item
    Optional domain mastery module: Additional minor research courses,
    but students with a home-domain already have the research part
    well-covered. Courses here should be allowed to fall in any research
    field, and those outside of the own home-domain should be especially
    encouraged.
  \item
    Research tools module: Here we teach tools used to distribute and
    publish software, as well as introducing students to domain specific
    data repositories, thereby gaining foundational knowledge in
    (\gls{SRU}, \gls{SP}, \gls{DOMREP}).
  \item
    Meta-research module: Here we teach people how research works. The
    research life cycle is introduced, as well as the data life cycle
    and the software life cycle are abstractly introduced.
  \end{itemize}
\item
  Communication skills:

  \begin{itemize}
  \tightlist
  \item
    Project management methods: Here we teach project management methods
    that are useful in science, such as agile ones (\gls{PM}).
  \item
    Communication skills module: Here we have courses focusing on
    interdisciplinary communication, interacting across cultures,
    communication in hierarchies, supporting end users effectively.
    These are all facets of the (\gls{USERS}) skill.
  \item
    Teaching module: This module covers topics to effectively design
    courses and teaching material for the various digital tools, thereby
    strengthening the (\gls{TEACH}) skill.
  \end{itemize}
\end{itemize}

Throughout the programme the values outlined in \cref{sec:values} are
incorporated into the sessions to raise awareness of the codes of
conduct and to put these values into ethical practice (see
e.g.~\autocite{Brown2024}).

Given that RSE work also involves a lot of craftsmanship skills,
hands-on practice is an integral part of the curriculum. At least two
lab projects are required within the mandatory curriculum. These should
be executed as a team and involve a question from a domain science. We
recommend covering both the candidate's home domain as well as a
different one. Ideally, projects stem from collaborations with
scientists within the institution and RSE students take the role of a
consultant. This setup strengthens the (\gls{TEAM}, \gls{TEACH},
\gls{USERS}) skill and encourages also the (\gls{MOD}) skill through
interaction.

To emphasise the exposure to domains outside their bachelor's degree
domain, we recommend that RSEs also support their non-home-domain
project with introductory courses from this discipline. This schools
their ability to quickly adapt their vocabulary and thinking to other
disciplines and is an aspect of (\gls{MOD}).

To align with the specialisations listed in this paper, example optional
modules include topics on \ac{HPC} engineering/parallel programming,
numerical mathematics/scientific computing, web technologies, data
stewardship, AI models/statistics, and community management/training.

The programme is finalised with a master's thesis which should be dual
supervised by an RSE supervisor from an actual project, and a domain
supervisor. The thesis should answer a relevant research question from
the domain using computational methods, strengthening (\gls{NEW}).
Software development is required, and the code is part of the gradable
deliverables. The RSE supervisor ensures and grades the software
craftsmanship aspects of the project. This setup ensures that we are
grading the effectiveness of applying RSE skills in an actual research
environment.

\subsection{An example of a possible career
path}\label{subsec:examplecareer}

\paragraph{Setting the stage}\label{setting-the-stage}

Meet Kay, Kim's~\autocite{Anzt2021} younger sister who currently studies
researchology in a bachelor's programme in the established domain of
researchonomy at University of Orithena (UofO). We will follow Kay's
fictional career to illustrate how education, job-experience and a
career in academic institutions could lead to become a successful RSE.
In Kay's world, some of the measures proposed in this paper have already
been implemented.

\paragraph{Bachelor's degree}\label{bachelors-degree}

Through a program like Software Carpentry~\autocite{CarpentriesSoftware}
or The Missing Semester~\autocite{Athalye2023}, Kay learns about using
computational tools to support the sophisticated statistical analysis
typical for researchology. She uses those tools to create and automate
the steps of processing data and producing outcomes for her bachelor's
thesis (generating plots with matplotlib and even CI for automatic
building) and takes pride in a fully open and reproducible bachelor's
thesis enabling her to graduate with honours from the faculty of
researchonomy.

\paragraph{Master's degree}\label{masters-degree}

Kay ponders whether to continue with computational researchology, which
her bachelor's supervisor is responsible for, or enrol in a
domain-agnostic RSE master's programme. Researchers in computational
researchology need to acquire a large part of the general RSE know-how
presented in this paper and specialise in Quantum-Accelerated Bayesian
Optimisation methods. However, Kay decides to go for the more generic
route of a dedicated RSE programme because she wants to continue in
academia, but does not like the idea of becoming stuck with one research
topic. She also experienced the immediate satisfaction gained by helping
colleagues from her research group with tricky technical problems, which
makes her happier than the subdued sense of achievement from having a
research paper accepted long after she had written it. For her, coding
and sharing knowledge in the form of software is of similar importance
to writing a paper focused mostly on the obtained results.

The domain-agnostic RSE Master programme consists of a core of RSE
topics with various electives for specialisation, some of them
domain-specific (e.g., chemistry) or topic-specific (e.g., cloud
computing for research). Kay chooses digital archaeology and develops a
pipeline for reconstructing 3D models from ground penetrating radar
data, to simplify the process for archaeologists (reproducibility, big
data, \ac{ML}). The project management skills that are being taught as
part of the core RSE curriculum really help her to not get lost in this
project. Apart from working with the researchers in her archaeology
group, she has to work with members of the central RSE department to
help her with the pipelines. She also has to liaise with the central
\ac{IT} department to organise storage for the large data sets. Towards
the end of the programme, she visits her first RSE conference where she
sees a lot of notions (\gls{SWLC}, \gls{RC}) in action that so far have
been abstract in her master's degree.

The exposure to the wider RSE community inspires her to invest
additional time into her thesis to publish her software project under a
licence approved by the Open Source Initiative and to write an
accompanying article in the open source journal JOSS~\autocite{JOSS}.
Inspired by the discussion with reviewers of her JOSS paper, and the
citation metadata file that JOSS created automatically for her when her
paper is published, Kay starts to think more about making her software
FAIR. She reads up on the topic in a guide suggested to her, the Turing
Way~\autocite{turing_way_2022}, and creates metadata files that provide
the citation metadata and general description for her software. She adds
the files to her source code repository, and also adds an automated
\gls{CI}/\gls{CD} pipeline that updates metadata and creates a new
publication record in the Zenodo repository for each new release. Kay
has now completed the RSE programme and has reached Junior RSE level.

\paragraph{Junior RSE}\label{junior-rse}

Kay finds a position in the central RSE department at her university
with a competitive \ac{IT} salary. Although the contract is temporary,
there is a good chance that it will lead to a permanent position. The
university makes an effort to enable that since it is a member of ``The
Technician Commitment''\autocite{techniciancommitment}, an initiative to
ensure recognition and career development of technicians, who face
similar challenges to RSEs. Kay completes the Software Carpentry
Instructor training and teaches basic research computing, while advising
fellow students of her department on better programming (\gls{DOCBB} and
\gls{MOD} skill). She also runs a seminar in the RSE Master's programme.
She publishes a condensed version of that in JOSE\autocite{JOSE}. During
her teaching duties, she becomes aware of a new project in her
department that requires a community manager RSE, and she gladly signs
up to focus more on her communication skills. After three years, she
takes an exciting opportunity to work in another university.

\paragraph{Senior RSE}\label{senior-rse}

The new position involves taking responsibility for the RSE related
aspects of a large inter-organisational project. With her new
responsibilities comes a shift in the importance of various aspects of
her work. Having this position in an inter-organisational project places
far more emphasis on communication and organisation skills. She is
spending time teaching people (\gls{TEACH} skill) to onboard them into
the project. There is a lot of interaction with different stakeholders
in the project like funders and user groups (\gls{USERS} skill). To
oversee the project, she uses an amalgamation of both agile and
traditional project-management concepts and methods which she acquires
on-the-job (\gls{PM} skills). Her work so far has already been heavy on
(\gls{TEAM}) skills, but now also the leadership aspect comes into play.

\paragraph{RSE-focused principal
investigator}\label{rse-focused-principal-investigator}

The job experience as a leading RSE for a large project was the last
requirement necessary to be awarded the title of a ``Certified Research
Software Professional'' (CRSP) from an institutionalised centre of RSE
education. The certificate confirms her track record of valuable
software contributions and of teaching and mentoring people, as well as
her capability to enable, foster and contribute to high-quality research
in a leading position. It is recognised by various funding agencies,
such as the DFG, and hence enables RSEs to act as a \ac{PI} for
RSE-focused grant applications. It is also recognised by many
prestigious universities and opens many career options that are also
typical for PhDs. Kay can now write her own grant proposals to
effectively fund work of moving research software projects from
prototypes to infrastructure.

\subsection{Existing frameworks}\label{subsec:existingframeworks}

\subsubsection{HPC skills and
certification}\label{hpc-skills-and-certification}

As an area that generally requires a range of advanced skills, \ac{HPC}
is one field where there is ongoing work to identify relevant sets of
skills for \ac{HPC} practitioners and potential paths to develop these
skills. The \acrshort{HPC} Certification
Forum~\autocite{HPCCFCompetencies} has developed a competence standard
for \ac{HPC} that defines a range of skills and how they are related in
the context of a skill tree~\autocites{Kunkel2020a}[~][]{Kunkel2020b}.
This competence standard is currently being built upon by the CASTIEL
2~\autocite{CASTIEL2} project in collaboration with initiatives funded
by the \ac{EuroHPC-JU} to create a framework for \ac{HPC}
certification~\autocite{EuroHPCJU2023}. While this framework focuses
mostly on skills specific to \ac{HPC}, there are a couple of
similarities to the framework proposed in this paper. The ``SD: Software
Development'' skill set is very similar to the \ac{SE} skills discussed
in \cref{sec:required-generic-skills}, describing a wide range of such
skills. This skill set contains Programming Best Practices (SD2),
Software Configuration Management (SD3), Software Quality (SD5),
Software Design and Software Architecture (SD6), and explicit mention of
documentation (SD7, see our \gls{DOCBB}). Besides the Software Concepts
for \ac{HPC} (SD1), which mainly concerns \acrshort{HPC}-focused RSEs,
most of the skills contained in the SD2-SD7 categories apply to all
RSEs. A significant difference compared to the framework proposed in
this paper is the absence of skills related to research or
communication. Noteworthy is already now the level of detail in their
skill tree which is more similar to \cref{subsec:examplemaster}.

Also looking at pathways and how different skills are related, the
\ac{UNIVERSE-HPC} project~\autocite{UNIVERSEHPC}, funded under the
\acrshort{UK}'s ExCALIBUR research programme~\autocite{EXCALIBUR}, is
looking to understand and develop training pathways to support the
development of specialist skills in the \ac{HPC} and exascale domains.
The project is gathering open source training materials to develop
curricula that support the training pathways that are underpinned by
high-quality training materials.

\subsubsection{Bioinformatics skills and
certification}\label{subsec:emblbio}

Bioinformatics is another field that actively works on developing skill
trees. The Bioinformatics Core
Competencies~\autocites{Mulder2018}[~][]{Welch2016}[~][]{Welch2014}, the
BioExcel competency framework~\autocite{Matser2016}, the PerMedCoE
competency framework~\autocite{Lloret-Llinares2021}, the Research Data
Management and Data Stewardship competence
framework~\autocite{Demchenko2021} and the ELIXIR Data Stewardship
Competency Framework for Life Sciences~\autocite{Scholtens2019} are
examples of grassroots efforts aiming at defining the set of skills of
various bioinformatics specialities, one of them as a
taxonomy~\autocite{Mulder2018}. These frameworks eventually converged
into the \ac{EMBL-EBI} Competency Hub
\autocite{CompetencyHub,Lloret-Llinares2022}, where typical RSE and
bioinformatician profiles at different levels of seniority can be
queried (e.g., Junior RSE\footnote{\url{https://competency.ebi.ac.uk/framework/bioexcel/3.0/profile/view/10115/alex-2}},
Senior Computational Chemist\footnote{\url{https://competency.ebi.ac.uk/framework/bioexcel/3.0/profile/view/10121/kim-0}})
and compared against one another (e.g., Junior vs.~Senior
RSE\footnote{\url{https://competency.ebi.ac.uk/framework/bioexcel/3.0/profiles/compare/10115/10117}}).

Competencies can be divided into more fine-grained building blocks:
knowledge, skills and abilities (KSAs). They can be organised in a
taxonomy, and are also transferable, i.e.~a KSA can be a prerequisite to
multiple competencies. The Mastery Rubric for
Bioinformatics~\autocite{Tractenberg2019} and the ELIXIR Data
Stewardship Competency Framework for Life
Sciences~\autocite{Scholtens2019} are examples of KSA frameworks for
bioinformatics curricula.

The Curriculum Task Force of the \ac{ISCB} curates a database of degrees
and certificates in
bioinformatics~\autocites{BioinformaticsCertification}[~][]{Mulder2018}.
The database includes bachelor's and master's degree programs and
specialisations, PhD programs, and certificates from graduate schools.

BioExcel has research competencies that combine some of our research
competencies and some notions from the communication skills. Their
computing competencies roughly map to our software skills. Here, we find
competencies such as ``package and distribute software'', which maps to
our (\gls{DIST}) competencies, and ``comply with licensing policy'',
which would in our framework be part of (\gls{SP}) in the research
competencies. In addition, they have a dedicated parallel computing
competency section, thereby shifting the emphasis of the knowledge of
their computational tools towards the \acrshort{HPC}-RSE specialisation
in our framework. Career profiles, such as the computational chemist,
bring additional domain specific knowledge; we would classify those as a
mixture of \$\{DOMAIN\}-RSE and \acrshort{HPC}-RSE. It is noteworthy,
however, that the BioExcel framework puts very little emphasis on
communication skills, which are often involved in RSE-related tasks.

\printbibliography

\printglossary

\printglossary[type=skills]

\printglossary[type=\acronymtype]

\end{document}